\documentclass[aps,prc,twocolumn,amsmath,floatfix,superscriptaddress,nofootinbib]{revtex4}
\usepackage{color,graphicx,ulem,subfig,soul,multirow}
\usepackage{mathptmx}                
\usepackage{dcolumn}                 
\usepackage{bm}                      
\usepackage{epsfig,amsmath}
\usepackage{xcolor}

\begin{document}
\title{Stellar electron capture rates on neutron-rich nuclei 
and their impact on core-collapse}

\author{Ad. R. Raduta}
\affiliation{IFIN-HH, Bucharest-Magurele, POB-MG6, Romania}
\author{F. Gulminelli}
\affiliation{LPC (CNRS/ENSICAEN/Universit\'e de Caen Basse Normandy), 
UMR6534, 14050 Caen c\'edex, France}
\author{M. Oertel}
\affiliation{LUTH, CNRS, Observatoire de Paris, Universit\'e Paris Diderot, 
5 place Jules Janssen, 92195 Meudon, France}

\begin{abstract}
During the late stages of gravitational core-collapse of massive
stars, extreme isospin asymmetries are reached within the core. Due to
the lack of microscopic calculations of electron capture (EC) rates
for all relevant nuclei, in general simple analytic
parameterizations are employed. We study here several extensions of
these parameterizations, allowing for a temperature, electron density
and isospin dependence as well as for odd-even effects. 
The latter extra degrees of freedom
considerably improve the agreement with large scale microscopic rate
calculations.  We find, in particular, that the isospin dependence
leads to a significant reduction of the global EC rates during core
collapse with respect to fiducial results, where rates optimized on
calculations of stable $fp$-shell nuclei are used.  Our results
indicate that systematic microscopic calculations and experimental
measurements in the $N\approx 50$ neutron rich region are desirable
for realistic simulations of the core-collapse.
\end{abstract}

\today

\maketitle

\section{Introduction}

Electron capture (EC) on nuclei plays a crucial role in high density
astrophysical environments and determines many processes such as
$r$-process nucleosynthesis \cite{Goriely15}, heating \cite{Gupta06} and
cooling of the accreting neutron star crust \cite{Schatz14}, thermonuclear
explosions of accreting white dwarfs and explosive nucleosynthesis
\cite{Iwamoto99,Brachwitz00}, the late stage evolution of massive stars
\cite{Aufderheide94,Heger2001} and core-collapse (CC) supernovae
\cite{Hix2003,Janka2007,Langanke03}.

In particular, during late stage stellar evolution EC is responsible
for a number of important phenomena.  First, it reduces the pressure
provided by the degenerate relativistic electron gas in order to
balance gravitation.  It also cools the core by producing neutrinos
which, for mass densities smaller than about $10^{11}$ g/cm$^3$ 
(which corresponds to baryon number densities, $n_B$, of about 
$6 \cdot 10^{-5}$ fm$^{-3}$) leave the star unhindered.  
Finally, EC determines the star's composition via the
electron-to-baryon ratio $Y_e$, equal to the proton fraction $Y_p =
n_p/n_B$ due to charge neutrality. $n_p$ denotes here the charge density. 

In a previous paper \cite{Raduta16} we have shown that nuclear
statistical equilibrium (NSE) averaged EC rates strongly depend, in
the late stages of pre-bounce evolution of the central element of a
collapsing star, on the experimentally unknown binding energies of
nuclei far beyond the stability valley.  By controlling the evolution
of $N=82$ and $N=50$ shell gaps in neutron-rich nuclei, we have shown
that nuclear abundances can be strongly modified. As a consequence,
the average EC rate may increase up to 30\%.  This study already
underlined the need of additional experimental measurements and/or
microscopic calculations at the neutron-rich edge of the known
isotopic table.

Total EC rates are not only influenced by the nuclear distribution,
but by the rates on individual nuclei, too.  This aspect was recently
addressed in Ref.~\cite{Sullivan16}.  Core collapse simulations were
performed, considering a huge variety of progenitor models and several
different equations of state (EoS).  Individual EC rates were thereby
systematically scaled by factors ranging from 2 to 10 with
respect to the fiducial values taken from
Refs.~\cite{LMP,Langanke03,Oda} and the analytic parameterization of
Ref.~\cite{Langanke03}.  
The mass of the inner core at bounce and the maximum of the neutrino
luminosity peak were found to augment (diminish) by 16\% (4\%) and, 
respectively, 20\% when the EC rates are scaled by a factor of 0.1 (10).
Moreover, it was shown that the
modification of EC rates affects much more strongly the bounce
properties than the progenitor model or the EoS. The results of the
simulations were found to be most sensitive to a possible reduction in
the EC rates for neutron rich nuclei near the $N$ = 50 closed
shell. This finding is very interesting for future experiments with
exotic beams.

This pedagogical study has clearly shown how much the lack of reliable
information on EC rates on the relevant neutron-rich nuclei can impact
the dynamics of core collapse.  However, a global modification of all
EC rates by a common factor -- the same for all nuclei and all
thermodynamic conditions -- is clearly not realistic.  Therefore,
awaiting for more detailed microscopic calculations, it is timely to
try to understand the different physical effects entering the EC rates
in order to better control them outside the region where calculations
constrained by experimental data are available.  To this aim, instead
of arbitrarily varying individual rates as in
Ref.\cite{Sullivan16}, we try here to obtain some hints on their
physical behavior at high electron density, temperature and isospin
ratio from the trends observed in the region covered by microscopic
calculations. Specifically we extend the existing analytic
parameterizations in several ways to incorporate different physical
effects and improve the reliability of the extrapolation to regions
not covered by microscopic calculations.

We shall show that these improved parameterizations can lead to a
reduction of the average EC rate with respect to previous
estimations. Over a certain density domain, the reduction factor is as
high as one order of magnitude, which corresponds to the maximum
reduction considered in Ref.~\cite{Sullivan16}.

To illustrate our findings, we will consider here some typical
thermodynamic conditions. They are taken from two different core
collapse trajectories reported in Ref. \cite{Juodagalvis_2010}.  They
correspond to the pre-bounce evolution of the central element of the
star with an enclosed mass of 0.05 solar masses, for two progenitors,
a $15 M_{\odot}$ and a $25 M_{\odot}$ one~\cite{Heger2001,Hix2003}.
Similar to our previous work, see Ref.~\cite{Raduta16}, the modified
EC rates will be added perturbatively.  This means that we neglect for
this exploratory work the influence of the modified rates on the time
evolution of baryon number density $n_B$, temperature $T$ and proton
fraction $Y_p$ throughout the collapse. Consequently, our
quantitative predictions contain some uncertainties, and we cannot
exclude that a consistent use of improved EC rates in a full core
collapse simulation would produce an effect different 
than the one shown in the present work.  
However, qualitatively, our results are robust
and the observed effect is large enough such that we expect in any
case that it will have non-negligible impact on the dynamics of the
collapse.

Given the thermodynamical conditions, we evaluate the chemical
composition of matter and in particular nuclear abundances via the extended NSE
model of Ref.~\cite{Gulminelli15}. Experimental nuclear masses
\cite{Audi} supplemented by predictions of the 10-parameter mass model
of Duflo and Zuker \cite{DZ10} are used.  EC rates are calculated
based on an analytic parameterization proposed in
Ref.~\cite{Langanke03}, fitted on large scale shell model calculations
of $fp$-shell nuclei~\cite{LMP,Langanke00}.  Three refinements of the
original parameterization will be proposed: The average Gamow-Teller
(GT) transition energy $\Delta E$ will be first allowed to depend on
temperature and electron density, second on isospin, and finally, on
the odd-even character of individual nuclei.  These latter
dependencies are inspired by the isospin asymmetry and odd-even
dependence of the GT$_+$-resonance, shown by both experimental data
and theoretical calculations (see for instance
\cite{Juodagalvis05,Langanke00,Batist10}).

The paper is organized as follows. Information on the thermodynamic
conditions of the considered core collapse trajectories and the
corresponding matter composition are given in
Section~\ref{sec:thermo}.  The different prescriptions for the
calculation of EC rates are discussed in
Section~\ref{sec:ecrates}. The resulting EC rates averaged over the
nuclear distributions obtained from the NSE along the considered
trajectories are presented in Section~\ref{sec:average}. In
Section~\ref{sec:conclusions} we summarize the present work.

\section{Thermodynamic conditions and matter composition} 
\label{sec:thermo}
\begin{figure}
\begin{center}
\includegraphics[angle=0, width=0.9\columnwidth]{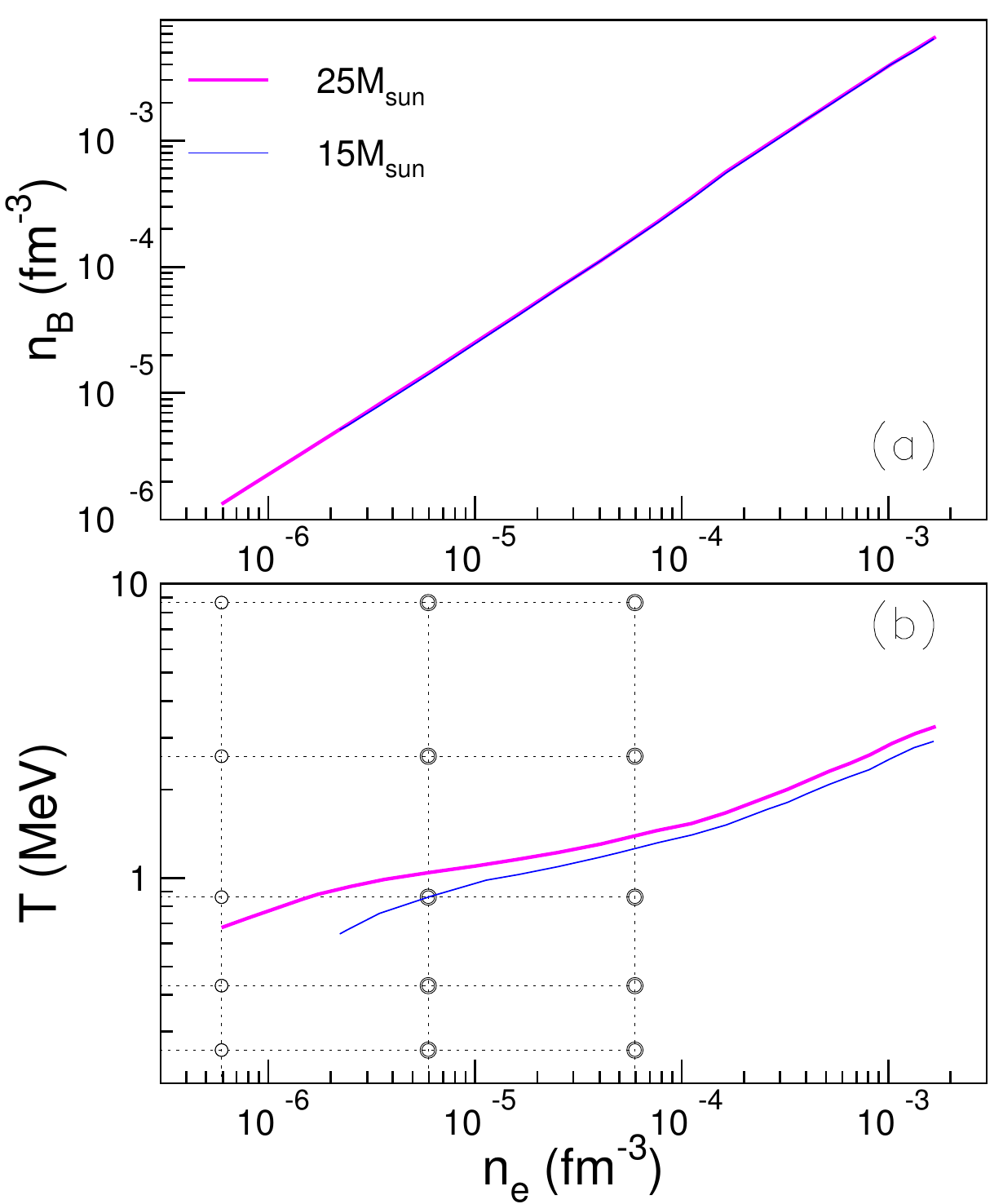}
\end{center}
\caption{Thermodynamic conditions explored during collapse by the
  central elements of two progenitors, a $15 M_{\odot}$ and $25
  M_{\odot}$ one, from Ref.\cite{Juodagalvis_2010}. Baryon
  number density (a) and temperature (b) are plotted as a function of the electron
  density $n_e$.  The grid on panel (b) indicates the
  temperature-electron density grid of the tables for weak interaction
  rates from Ref.~\cite{LMP}.
\label{fig:thermo}}
\end{figure}

\begin{figure}
\begin{center}
\includegraphics[angle=0, width=0.9\columnwidth]{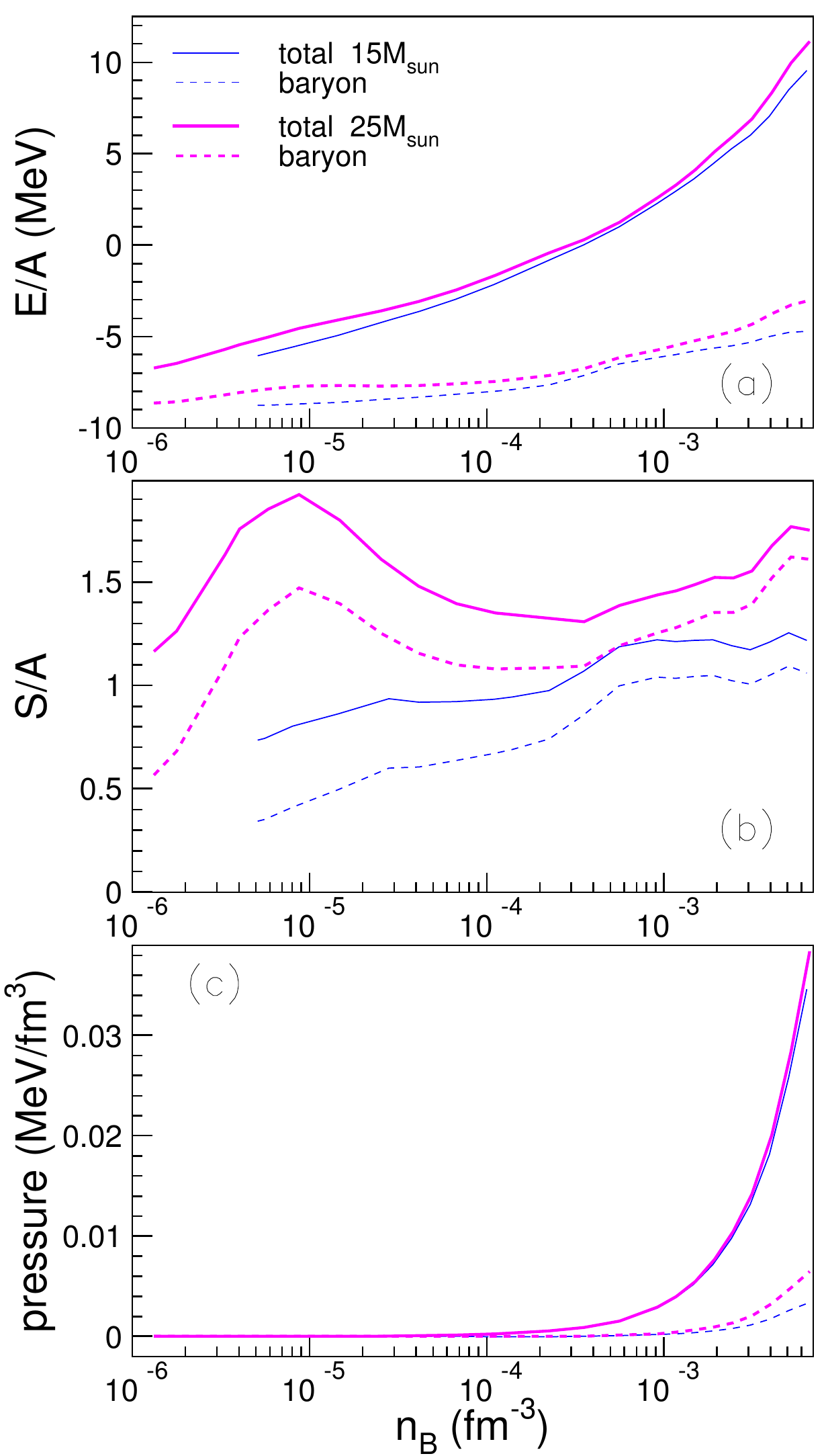}
\end{center}
\caption{Results of the extended NSE model of Ref.~\cite{Gulminelli15}
  using thermodynamic conditions reported in
  Ref. \cite{Juodagalvis_2010} of Fig. \ref{fig:thermo}.  Baryonic and
  total energy per baryon (a), entropy per baryon (b) and pressure (c)
  are plotted as a function of baryonic density. }
\label{fig:NSEresults}
\end{figure}

The pre-bounce evolution of a collapsing star follows a complicated
trajectory in the $T$, $n_B$ and $Y_p=Y_e$ space in response to different
physical processes, including weak interactions.  
However, according to the theory of adiabatic collapse \cite{Goldreich}, 
the central part of the core is expected to collapse homologously, 
meaning that the central densities and temperatures should manifest a 
more or less universal pattern taken as function of $Y_e$, 
as confirmed by detailed numerical simulations \cite{Yahil83,Bruenn85,Aufderheide94}. 

In Fig.~\ref{fig:thermo}, we show the thermodynamic conditions at the
center of the star during infall, resulting from two
simulations~\cite{Heger2001,Hix2003} with two different progenitors, a
15 $M_\odot$ and a 25 $M_\odot$ one. The central element within these
spherically symmetric simulations corresponds to an enclosed mass of
0.05 $M_\odot$. Weak interactions rates have thereby been determined
from a simple NSE matter composition together with rates taken
from Refs.~\cite{Langanke00,LMP}, see Ref.~\cite{Juodagalvis_2010}.
On panel (b), the electron density-temperature grid of the weak
interaction data base of Ref.~\cite{LMP} is represented as well. It is
obvious that the highest electron densities reached during infall
lie outside the grid domain and the temperatures, $T$, are situated in
a region where the mesh is sparse.  

Fig.~\ref{fig:NSEresults} offers complementary information on the considered trajectories.  
It shows, as a function of the baryon number density, the baryonic and total energy per baryon
(panel (a)), entropy per baryon (panel (b)) and pressure (panel (c))
as obtained within the extended NSE model of Ref.~\cite{Gulminelli15}.  We can see that
the slight progenitor dependence exhibited by the temperature
evolution (Fig. \ref{fig:thermo}(b)) is somewhat amplified in other thermodynamic
variables, notably the entropy per particle.
This is due to the fact that these quantities strongly depend on the
matter composition, which depends exponentially on temperature.  Thus,
even if the inner core evolves homologously, some progenitor
dependence remains within the thermodynamic conditions and we will
keep thus both trajectories for the discussion within this paper.
 
The matter composition along the two trajectories was investigated in
detail in Ref.~\cite{Raduta16}.  It was shown in particular that the
distributions of nuclei are broad at all times, and that individual
nuclei become more neutron rich subsequently to the deleptonisation
and decrease in $Y_p$ of the whole system.  Nuclei with a mass number
$A > 20$ (called "heavy" hereafter) bind an important amount of
matter, even if their mass fraction decreases as the collapse advances
and the temperature rises.  In addition, if binding energies of
nuclei are described by a model with strong shell gaps far from
stability, as the DZ10 model~\cite{DZ10}, the nuclear distributions are
characterized by a competition between the $N=50$ and $N=82$ magic
numbers \cite{Raduta16}.

\section{Electron capture rates}
\label{sec:ecrates}
The first systematic study of weak interaction rates under stellar
conditions is due to Fuller, Fowler and Newman. In a series of papers
published in the '80s, they pointed out the particular role played by
the GT giant resonance for both electron capture and $\beta$-decays,
and parameterized its contribution based on an independent particle
model \cite{FFN80-82,FFN82,FFN85}.  The resulting reaction rates for
nuclei with mass numbers between 21 and 60, considered of chief
importance in astrophysics, were made available in tabulated
form~\cite{FFN82}. Analytic parameterizations, where structure and
phase space contributions are factorized, were proposed in
Ref.~\cite{FFN82}, too. 

It is well known since these pioneering works that capture rates on
heavy nuclei dominate the capture probability, though lighter nuclei
cannot be neglected \cite{Raduta16,Fuller10}.  This is due to the fact
that the probability of capturing an electron is trivially hindered
for large and negative reaction Q-value, defined as
$Q(A,Z)=M(A,Z)-M(A,Z-1)+m_e$. $M(A,Z)$ denotes here the mass of a
nucleus with $Z$ protons and $N = A-Z$ neutrons, and $m_e$ is the
electron mass. The $Q$-value gives the relevant energy scale of the
capture process at constant electron energy. Light nuclei, for which few stable
isobars exist, have in general very low $Q$-values. For $fp$-shell nuclei
around $A\approx 50$, the nuclear chart starts to widen and the
$Q$-value increases.

\begin{figure*}
\begin{center}
\includegraphics[angle=0, width=0.8\textwidth]{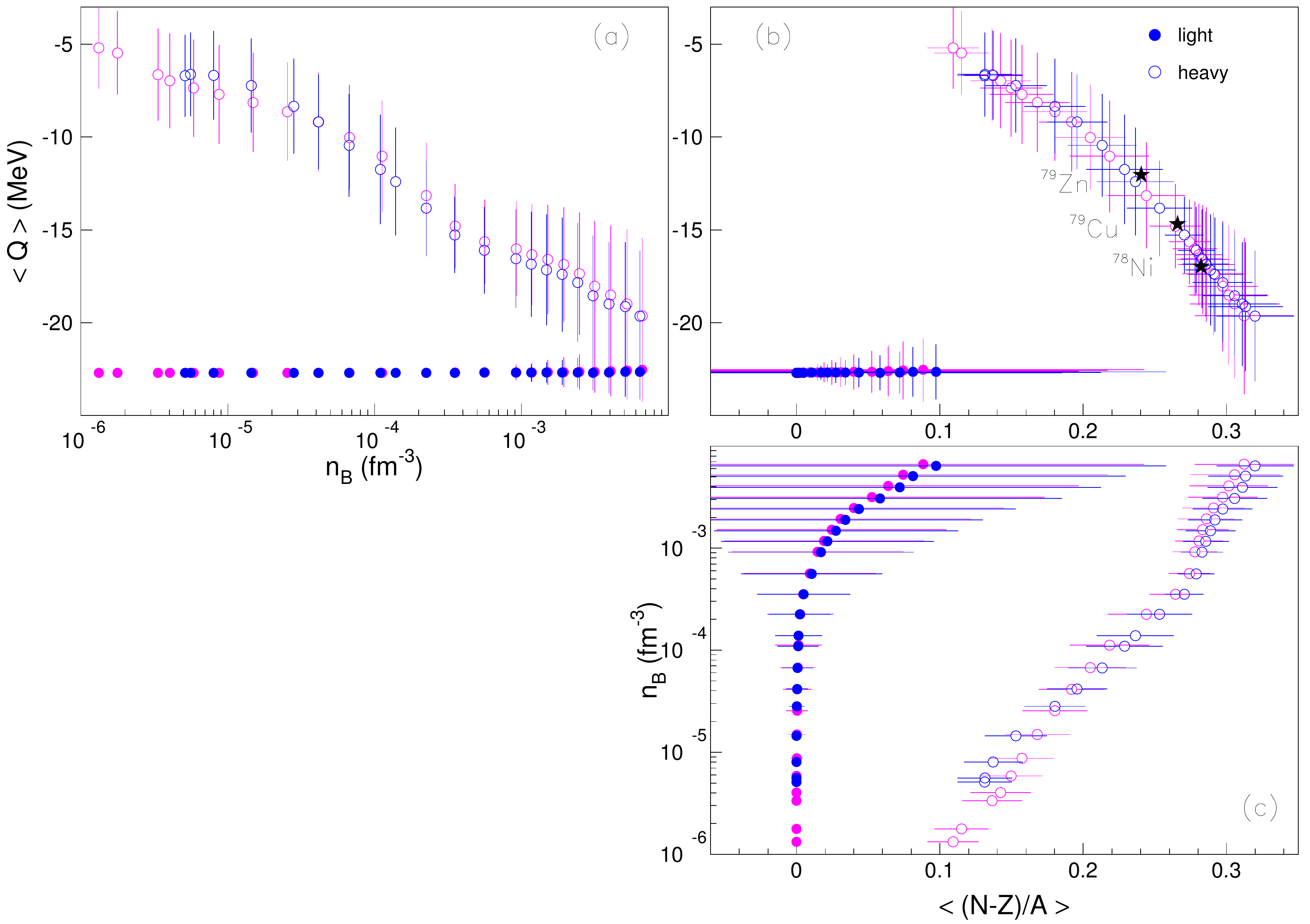}
\end{center}
\caption{Average electron-capture $\langle Q\rangle$-value 
as a function of baryon number density for
  heavy ($A \geq 20$) and light ($2 \leq A <20$) clusters (a), the same 
versus
  average isospin asymmetry $\langle  (N-Z) /A \rangle$ (b) and
  $\langle (N-Z) /A \rangle$ versus baryon number density (c).  
  The thermodynamic conditions correspond to the
  central element of the core during collapse, taken from
  Ref.~\cite{Juodagalvis_2010}, and the nuclear abundances are
  calculated within the extended NSE model of Ref.~\cite{Gulminelli15}.
  Vertical and horizontal error bars correspond to the standard deviations of 
  the $Q$ and $I$ distributions.  The color legend is the same as in
  Fig.~\ref{fig:thermo}. Solid stars on panel (b) mark the nuclei 
  which have been identified in Ref.~\cite{Sullivan16} as playing, from their
  individual rates, the dominant role for electron capture in CCSN
  simulations. }
\label{fig:IQ}
\end{figure*}

Of course, lower $Q$-values might be compensated by higher abundances
such that in an astrophysical situation still light nuclei may dominate. 
Therefore, we show in Fig.~\ref{fig:IQ}, the average $Q$-value, 
$\langle Q \rangle=\sum_{A,Z} n_{A,Z} Q(A,Z)/\sum_{A,Z}n_{A,Z}$, 
and average isospin asymmetry, 
$\langle I \rangle=\sum_{A,Z} n_{A,Z} (1-2 Z/A)/\sum_{A,Z} n_{A,Z}$,
throughout the CCSN trajectories of Ref.~\cite{Juodagalvis_2010}.
$n_{A,Z}$ represents the abundance per unit volume of
the nucleus with $A$ nucleons and $Z$ protons.
Nuclear abundances have been calculated within the extended NSE model of
Ref.~\cite{Gulminelli15} as detailed in the previous section. Light
nuclei ($2 \leq A<20$) are labeled by filled dots and heavy nuclei
($A \geq 20$) by open dots. The errors bars indicate the width of the
nuclear distributions.  It is obvious that for the considered
thermodynamic conditions within the collapse phase of a CCSN, heavy
nuclei systematically have larger average $Q$-values and, 
therefore, higher EC rates.

The widening of the region of nuclear stability also implies that
higher values of the average isospin asymmetry  can be
reached for heavier nuclei.  As expected from elementary
considerations on nuclear stability, and clearly visible on
Fig.~\ref{fig:IQ}, $\langle Q \rangle$ decreases with increasing
neutron richness. Furthermore, due to the competition among $N$-magic numbers
\cite{Raduta16}, wide distributions characterize heavy fragment
production under all thermodynamic conditions.  
Contrary to heavy clusters whose properties vary strongly as the collapse
proceeds,
light clusters, defined on a quite limited mass domain and highly dominated 
by $\alpha$-particles, show an almost constant $\langle Q \rangle$-value and, 
in the late stages before bounce, a very large variation with isospin.
The latter effect arises from the increasing contribution of 
loosely bound nuclei with $Z>2$ and large isospin,
allowed by high temperatures and densities and low $Y_p = Y_e$ values,
met in the late stage of the collapse.
 
The evaluation of EC rates for heavy nuclei demands accurate
microscopic calculations beyond the simple independent particle
picture employed in the first theoretical works.  Indeed, a clear
experimental evidence exists that the GT strength is quenched with
respect to single particle expectations, and fragmented over several
states at low excitation energies in the daughter nucleus.  The best
theoretical description presently available is offered by large scale
shell model calculations, able to account for all correlations among
the valence nucleons in a major shell.  So far such tabulations exist
only for $sd$- ($17 \leq A \leq 39$) \cite{Oda,Martinez2014} and
$fp$-shell nuclei ($45 \leq A \leq 65$) \cite{LMP}, for the same
temperature-electron density grid previously considered in
Refs.~\cite{FFN82}. A full diagonalization of the shell model
Hamiltonian is not yet feasible for open shell neutron-rich nuclei
with $A>65$, which dominate electron capture for densities larger than
$10^{10}$ g/cm$^3$ \cite{Juodagalvis_2010}. Calculations exist
employing alternative approaches, such as RPA \cite{Nabi,Fantina} or
QRPA \cite{QRPA}. In general, these are less accurate than shell-model
calculations~\cite{Cole} since it is not clear whether all
correlations are correctly included.  An interesting and accurate
hybrid approach has been proposed in Ref.~\cite{Langanke03}, where RPA
calculations have been performed with occupation numbers taken from
shell model Monte-Carlo calculations.

Globally, theoretical evaluations of electron capture rates, resumed
in the available microscopic~\cite{Oda,LMP} or empirical~\cite{Pruet}
databases, do, however, neither cover all necessary nuclei nor are the
grids in temperature and electron density dense and extended enough
for a complete description of EC during core collapse.  For this
reason, the use of analytic parameterizations appears as an appealing
compromise \cite{Langanke03,Sullivan16}.  

The most popular analytic parameterization used in core collapse
simulations was proposed in Ref.~\cite{Langanke03} and reads
\begin{eqnarray}
\lambda_{EC}=\frac{\ln 2 \cdot {\mathcal B}}{K} \left(\frac{T}{m_e c^2} \right)^5 
\left[ F_4(\eta)-2 \chi F_3(\eta)+\chi^2 F_2(\eta)
\right]~.
\label{eq:lEC}
\end{eqnarray}
In this expression, $\chi=(Q-\Delta E)/T$, $\eta=\chi+\mu_e/T$, and
$m_e$ and $\mu_e$ stand for electron rest mass and chemical potential,
respectively. $F_i(\eta)$ denotes the relativistic Fermi integral,
$F_i(\eta)=\int_0^{\infty} dx x^k/(1+\exp(x-\eta))$.  $K=6146$ s is a
constant\cite{Hardy}, while ${\mathcal B}$ and $\Delta E$ represent an
average value for the (Gamow-Teller plus forbidden) matrix element,
and for the transition energy, respectively.  The constant values
proposed in Ref.~\cite{Langanke03}, ${\mathcal B}=4.6$ and $\Delta
E=2.5$ MeV, are shown to give a good overall qualitative reproduction
of the microscopically calculated EC rates for the thermodynamic
conditions explored by the central element before bounce
\cite{Langanke03}. 

However, the microscopic calculations show a large dispersion, and
deviations in individual rates of one order of magnitude or more are
observed with respect to this simple prescription. In addition, only a
limited number of nuclei is included in the adjustment, and therefore
in practice the analytic parameterization in Eq.~(\ref{eq:lEC}) is
extrapolated to much smaller $Q$-values, needed for neutron rich
nuclei, than those present in the fit. 

In Ref.~\cite{Sullivan16} it has been shown that a modification
in the EC rates for some key nuclei can have a considerable impact on
core collapse. Therefore, here we try to understand which is the major
physical effect neglected in the simple parameterization,
Eq.~(\ref{eq:lEC}), susceptible to improve the description of individual
EC rates. This study could on the one hand lead to a more reliable
parameterization based on existing microscopic data and on the other
hand motivate microscopic calculations to put the
parameterizations on a more firm ground.

To that end we will study four different generalizations of
Eq.~(\ref{eq:lEC}), hereafter called model (0), with parameters adjusted to microscopic rate
calculations.  To avoid discontinuities due to different theoretical
treatments, we have not mixed different data sets but rather
concentrated on the calculations from Ref.~\cite{LMP} as reference. We
have checked that the inclusion of the data set from Ref.~\cite{Oda}
only marginally changes our fit parameters. Details of the different
generalized parameterizations are given in appendix~\ref{app:fits}.

In order to have a correct description of the EC rate, ${\mathcal B}$
and $\Delta E$ should in principle depend on the thermodynamic
conditions, because of the increasing importance of excited states for
increasing temperature and electron energy. Therefore in the first
model, denoted model (1), we allow $\Delta E$ to depend on temperature
and electron density\footnote{Since ${\mathcal B}$ gives only an
  overall normalization and does not change the functional dependence
  of the rate, we keep the fiducial value ${\mathcal B}=4.6$.}. We
concentrate here on temperatures of the order of MeV and the highest
electron density values of the tables from Ref.~\cite{LMP}, 
since they are the most
relevant for the core collapse trajectories we are interested in.
\begin{figure*}
\begin{center}
\includegraphics[angle=0, width=0.8\textwidth]{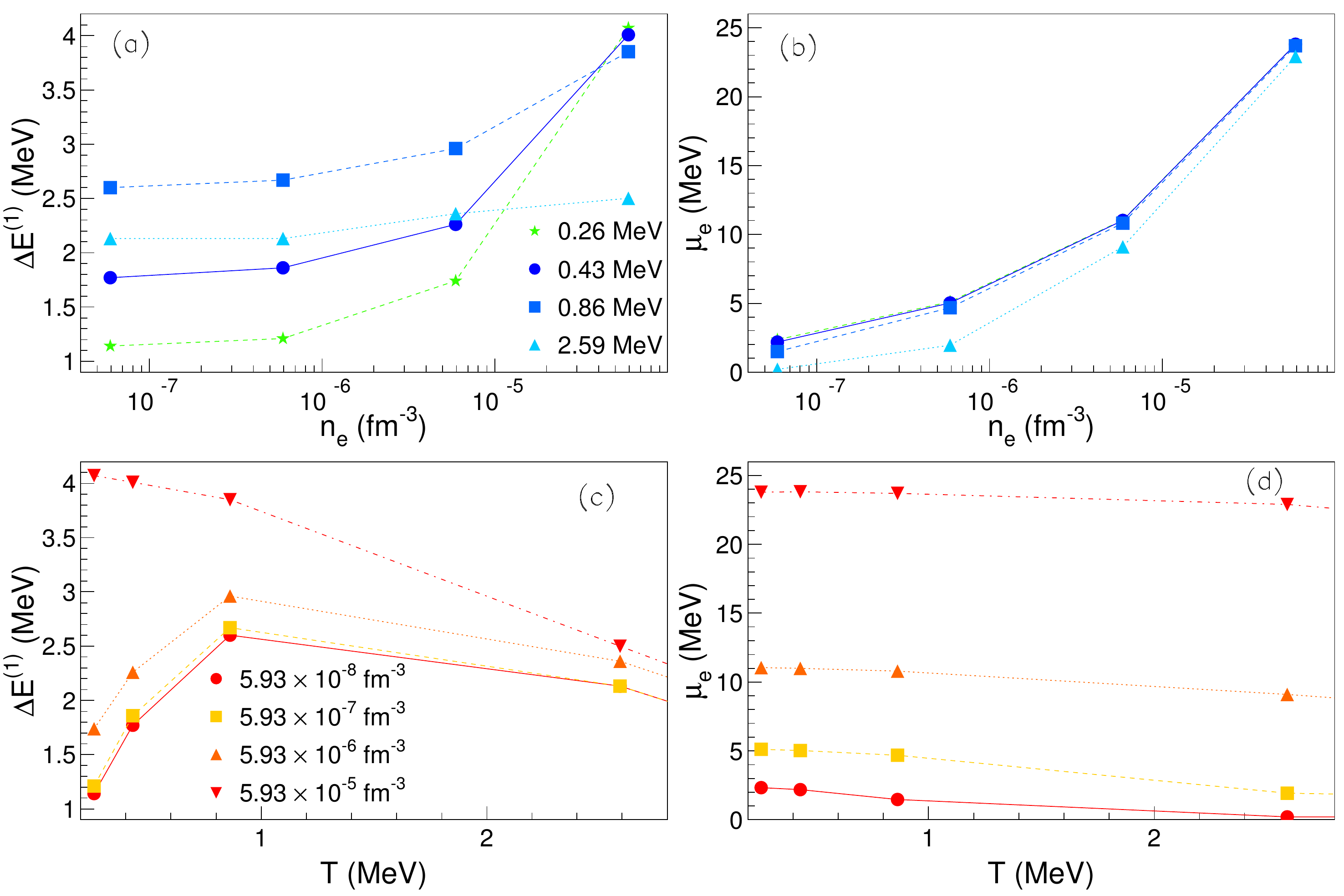}
\end{center}
\caption{Evolution of $\Delta E^{(1)}$ as a function of $n_e$ (a) and
  $T$ (c) for constant values of $T$ and, respectively, $n_e$. Panels
  (b) and (d) illustrate the evolution of electron chemical potential
  $\mu_e$ under the same conditions as in panels (a) and (c). 
\label{fig:DE_versusTne}}
\end{figure*}
In Fig.~\ref{fig:DE_versusTne} (left panels) we show the evolution we
obtain of $\Delta E^{(1)}$ with temperature and electron density,
respectively. 
The fitted values lie between 1 and 4 MeV for
all electron densities considered,
indicating that the simple choice of a constant $\Delta E = 2.5$ MeV
in Ref.~\cite{Langanke03} is a good first approximation. 

Looking more in detail, the behavior of $\Delta E^{(1)}$ as a
  function of temperature (panel (c)) becomes more complex. This can
be attributed to two competing effects. With increasing temperature,
the contribution of excited states in the daughter nucleus increases,
leading to a larger $\Delta E$. This effect is partly compensated by
an electron chemical potential $\mu_e$ decreasing with temperature,
see panel (d). Thus, depending on the value of $n_e$, $\Delta E^{(1)}$
can first increase and then decrease, or even show a monotonically
decreasing behavior throughout the entire considered temperature
range. 

The increase of $\Delta E^{(1)}$ with $n_e$, see panel (a) of
Fig.~\ref{fig:DE_versusTne}, can be attributed to the increase of
$\mu_e$ with $n_e$, allowing for more excited states with higher
energies to be populated. This effect is suppressed at the highest
temperature by the decrease of $\mu_e$ with temperature, see
above.

\begin{figure}
\begin{center}
\includegraphics[angle=0, width=0.99\columnwidth]{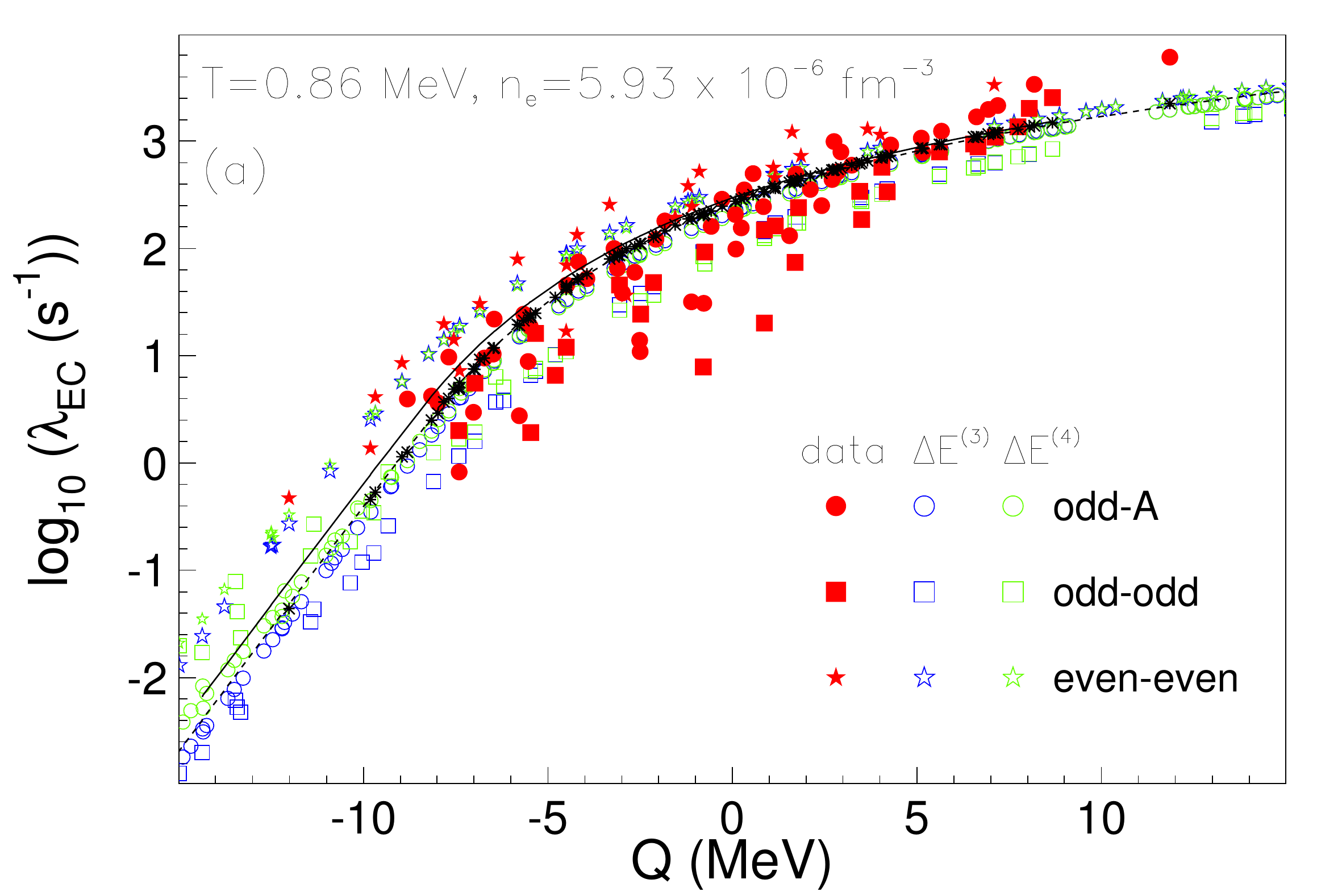}
\includegraphics[angle=0, width=0.99\columnwidth]{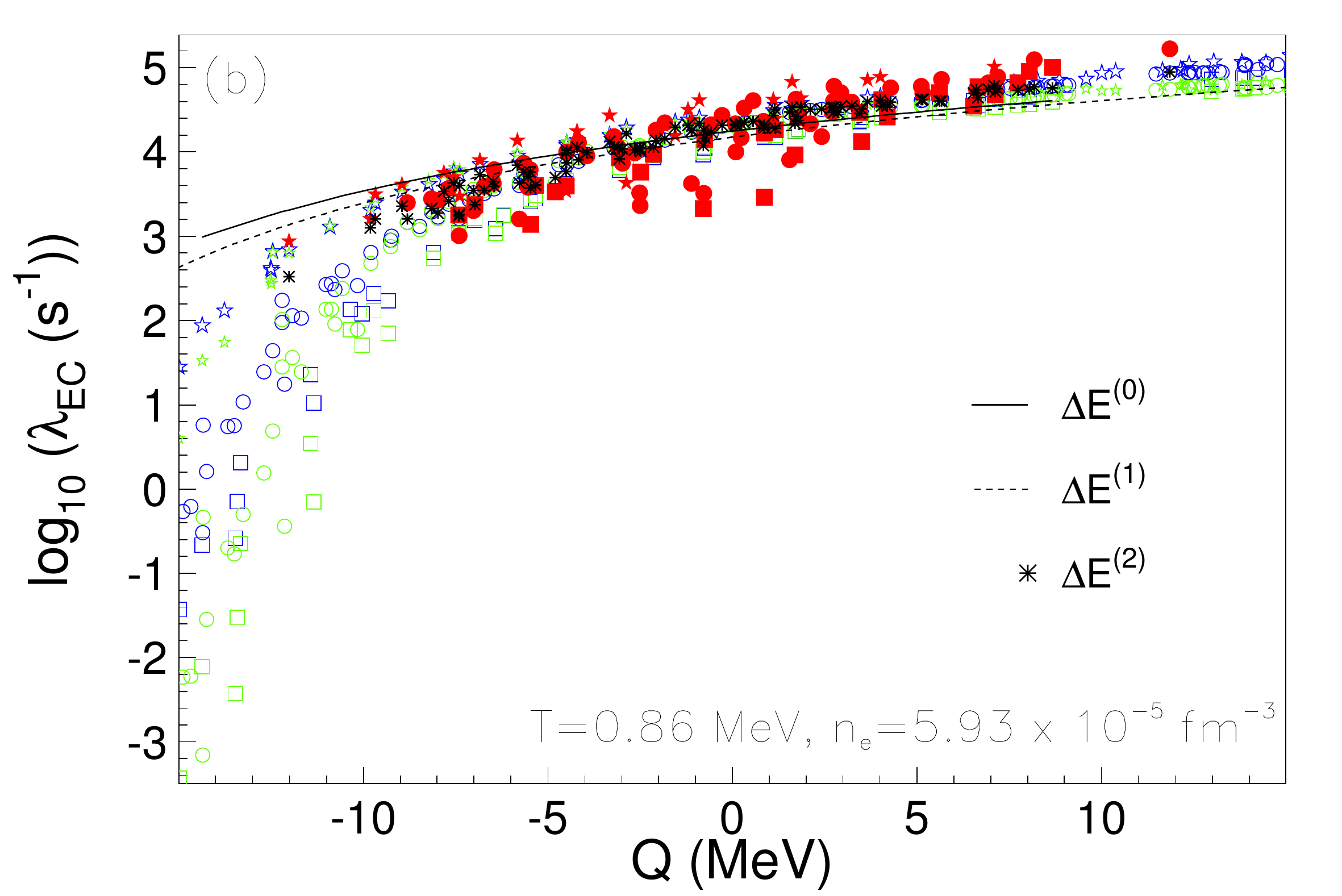}
\includegraphics[angle=0, width=0.99\columnwidth]{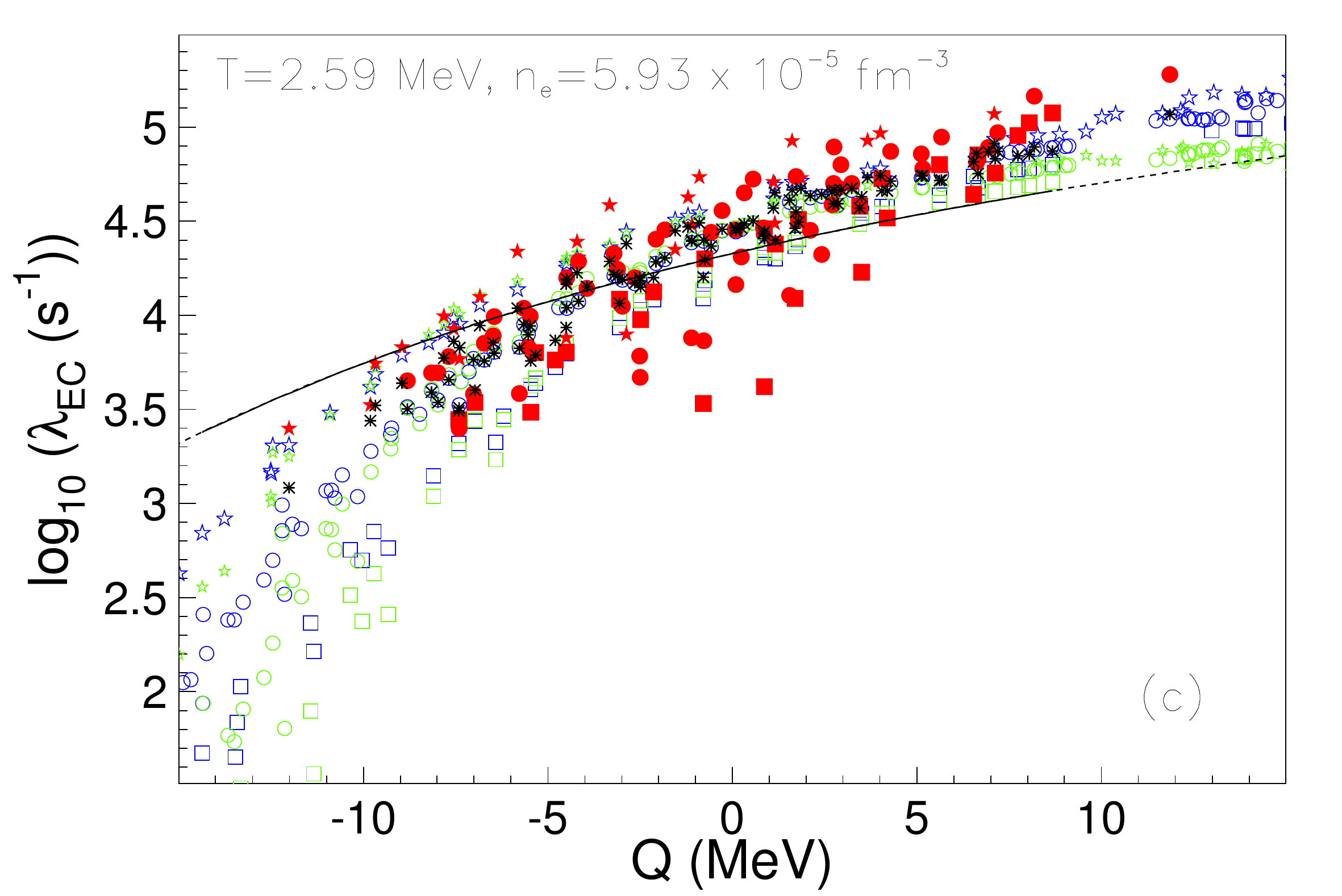}
\end{center}
\caption{Electron capture rates: comparison between predictions of
  Eq. (\ref{eq:lEC}) with different prescriptions for the average
  transition energy $\Delta E$ (see text) and values of the tables
  from Ref.~\cite{LMP}.  $\Delta E^{(0)}$ stands for the original parameterization of Ref. \cite{Langanke03}.
  The thermodynamic conditions, chosen from the grid points of weak interaction rates of Ref. \cite{LMP}, 
  are mentioned on each panel.
  The symbol and line legend is the same on the three panels.}
  \label{fig:EC}
\end{figure}

\begin{figure}
\begin{center}
\includegraphics[angle=0, width=0.99\columnwidth]{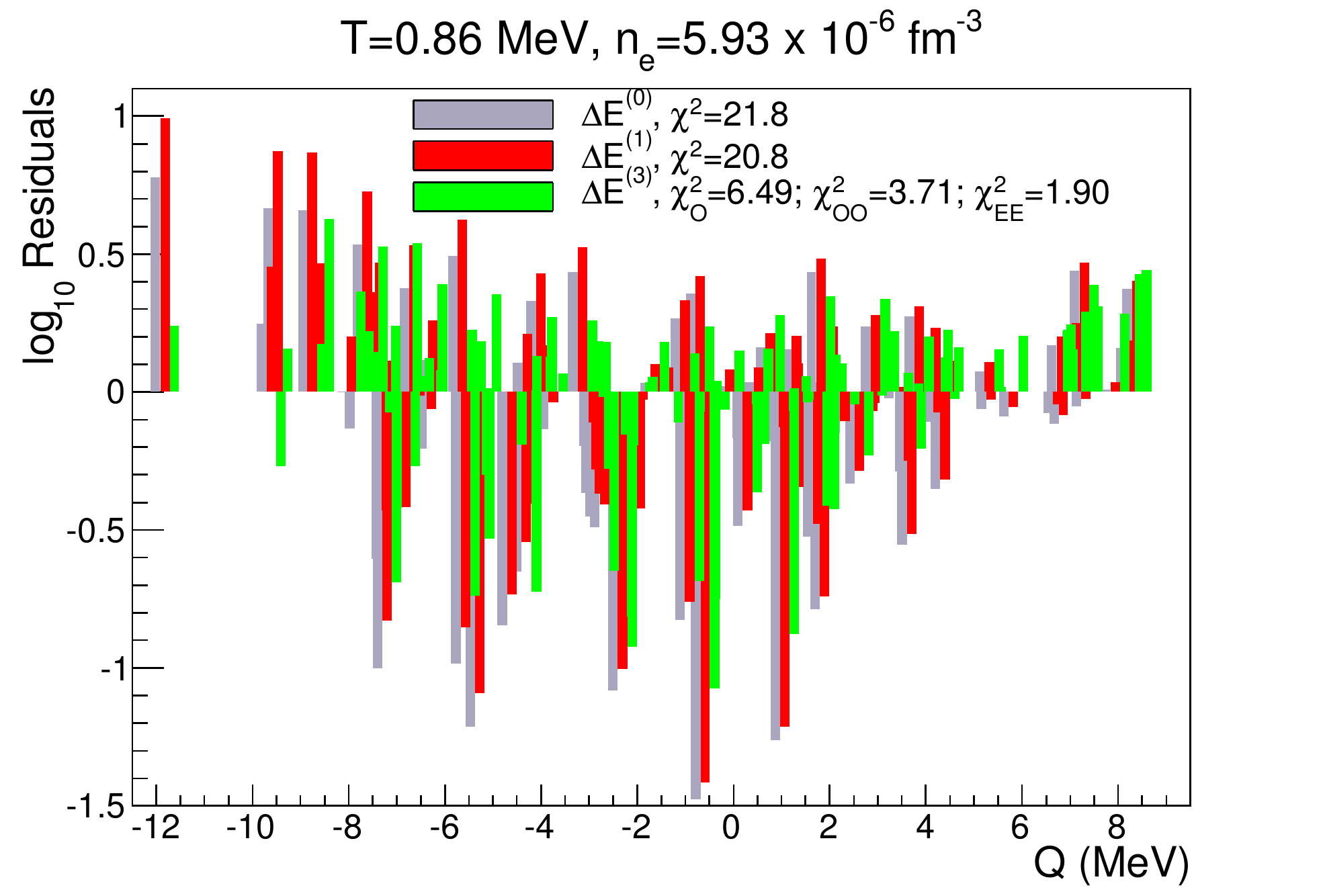}
\includegraphics[angle=0, width=0.99\columnwidth]{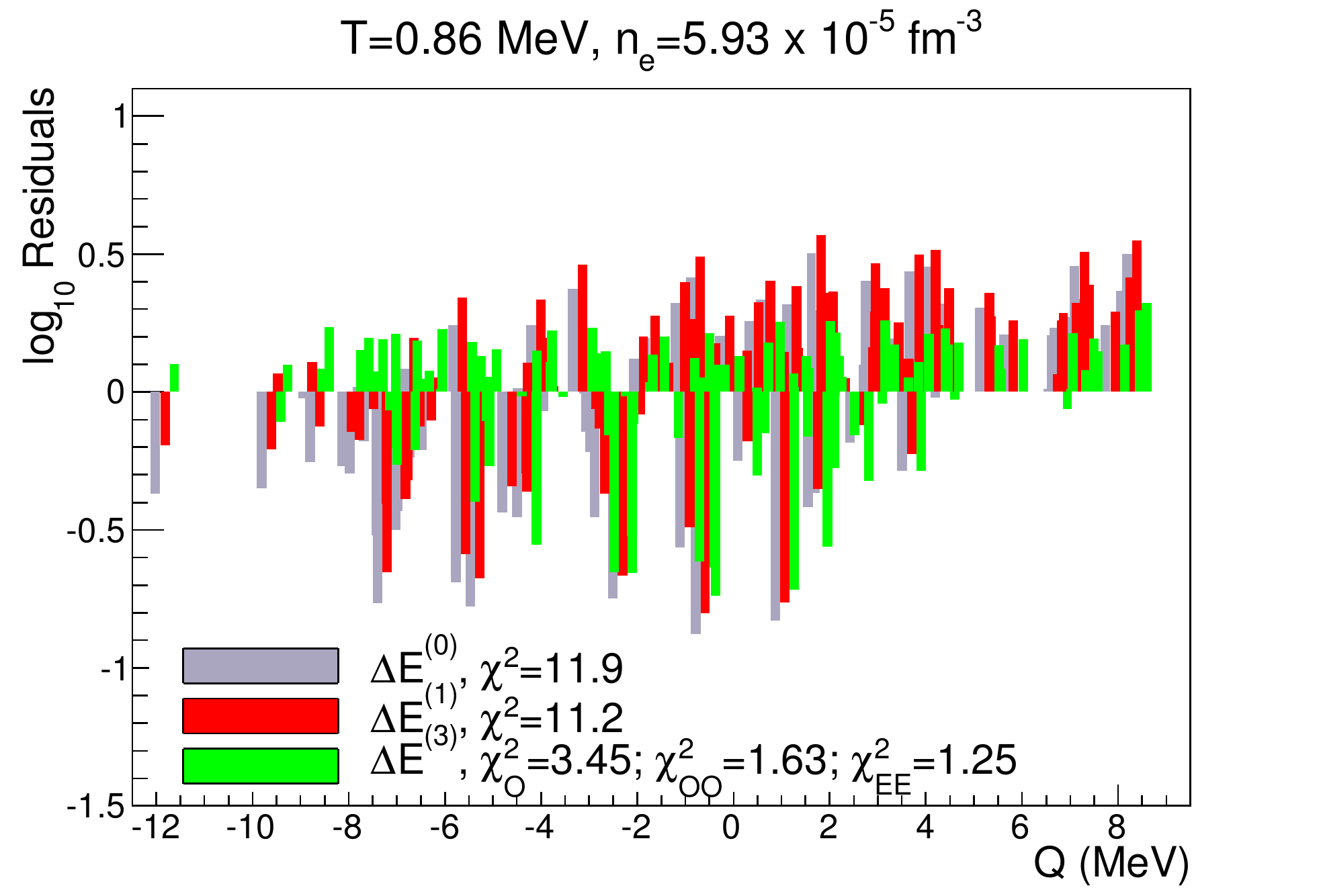}
\includegraphics[angle=0, width=0.99\columnwidth]{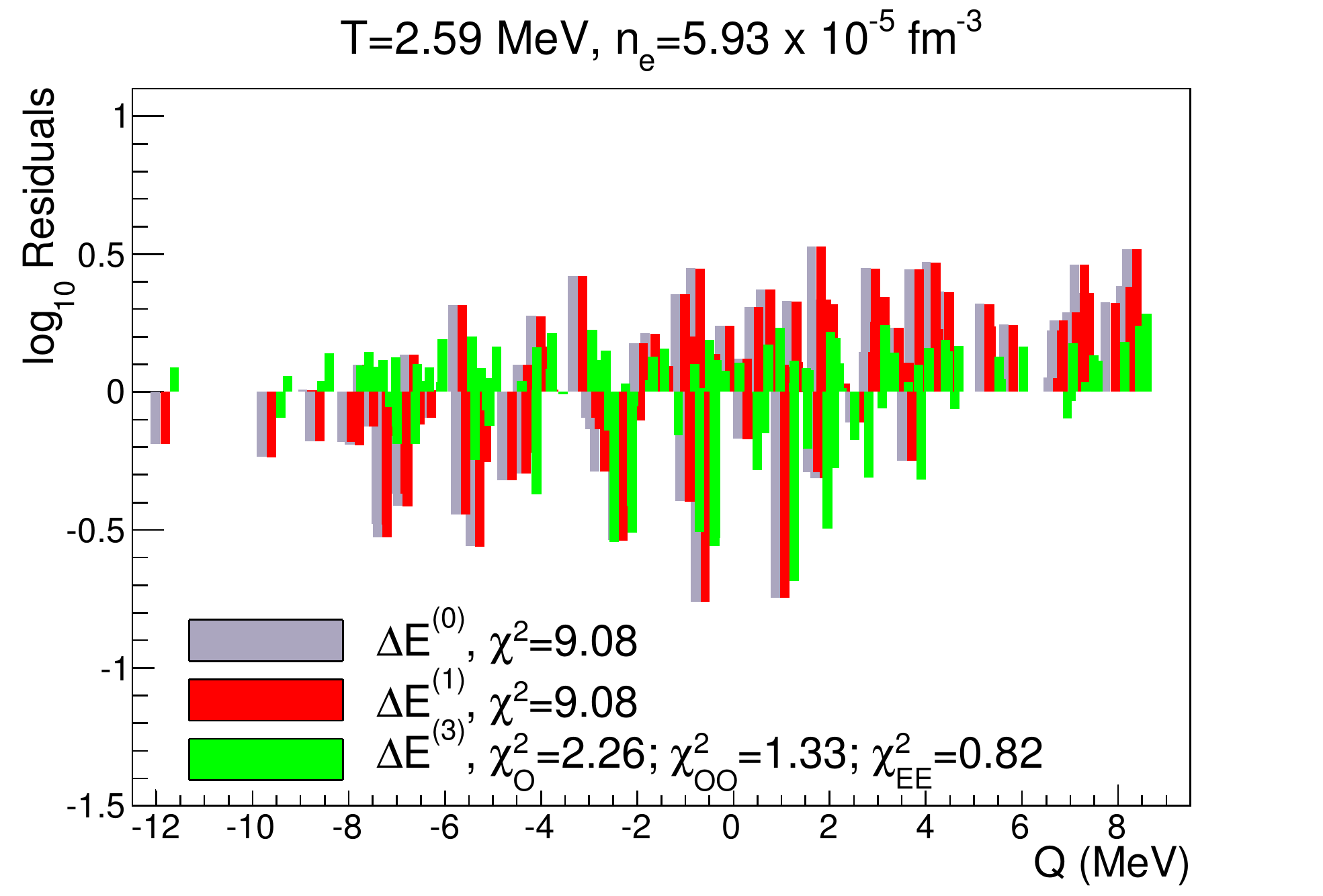}
\end{center}
\caption{The residual differences between $\log_{10}$ of the shell
  model rates of Ref. \cite{LMP} and the approximate rates for each
  nucleus in the weak interaction library calculated according to
  different recipes. The thermodynamical conditions, specified on each panel, 
  are the same as in Fig. \ref{fig:EC}. 
  To increase readability, for models (1) and (3) the $Q$-values have been 
  shifted by 0.2 and, respectively, 0.4 MeV. 
  In all cases, the $\chi^2$ values are mentioned in the key legend (see also Table I). 
\label{fig:residuals}}
\end{figure}

Altogether, allowing $\Delta E$ to depend on temperature and electron
density only marginally improves the overall agreement of the
parameterized rates with the microscopic ones, see Figs.~\ref{fig:EC}
and \ref{fig:residuals}. Fig.~\ref{fig:EC} displays a comparison of
the EC rates from the calculations of Ref.~\cite{LMP} with the
different parameterizations discussed here for several thermodynamic
conditions. The plain lines corresponds to Eq.~(\ref{eq:lEC}) and
parameters from Ref. \cite{Langanke03} (model (0)) and the dashed line
to model (1). In Fig.~\ref{fig:residuals} the residual differences are
shown. These simple parameterizations can clearly not reproduce the
large scattering of the shell model rates due to the density of states
of the daughter nucleus and the details of the GT strength
distribution. Further refinements are thus necessary for a good
description of EC rates.

In order to proceed, let us first note that the higher
$\mu_e$, the less important become the details of nuclear
structure. Therefore scattering of the rates is reduced with
increasing $n_e$ and the parameterizations better reproduce the data
(panels (a) and (b)). A similar effect occurs with increasing
temperature, since nuclear structure effects are partially washed out
by the increasing number of contributing excited states (panels (b) and (c)).  
It can be seen from Fig.~\ref{fig:EC}, too, that the
dependence of the EC rates on $Q$ considerably flattens with
increasing temperature and electron density.  As already observed in
\cite{Langanke03}, this is due to the fact that the electron chemical
potential increases with density much faster than the range of
$Q$-values explored by the most abundant nuclei, again reducing the
dependence on the detailed strength distribution. The capture process
is thus dominated by the centroid of the GT-resonance.  

Second, note that for high $T$ and $n_e$ values, a
systematic deviation of the parameterizations with respect to the data
of Ref.~\cite{LMP} can be observed (see Fig.~\ref{fig:EC}, panels
(b,c)): too high (low) EC rates for negative (positive) $Q$-values.  A
possible explanation for this deviation could be a residual isospin,
$I = (N-Z)/A$, dependence of $\Delta E$ of the centroid of the GT resonance~\cite{Juodagalvis05,Langanke00,Batist10}.

For this reason, in model (2) we further allow for a linear
$I=(N-Z)/A$-dependence of $\Delta E$, see appendix~\ref{app:fits} for
details. The overall agreement of the parameterization with the data
is considerably improved, see Fig.~\ref{fig:EC} and
Table~\ref{table:chi2}, even if at low temperatures noticeable
deviations remain. This is due to the fact that the individual rates
under these conditions depend on the details of the GT strength and it
is thus not sufficient to reproduce global trends of the GT
phenomenology.

A closer look at Fig.\ref{fig:EC} reveals that the dispersion of the
individual rates is partially due to the existence of strong odd-even
effects, 
see also the discussion in Ref.~\cite{Langanke00}.  We
therefore introduce a separate parameterization of the average transition energy
parameter $\Delta E$ for odd-odd (OO), even-even (EE) and odd-even
(OE) nuclei, called model (3), see appendix~\ref{app:fits} for details.

This extra refinement proves to be crucial for a good reproduction of
shell-model calculations.  As we can see from Table~\ref{table:chi2}
(columns 11-13) and Figs.~\ref{fig:EC}, \ref{fig:residuals}, a sizeable
reduction of the residuals is observed and data are well
reproduced under all thermodynamic conditions considered here. Quite
interesting, the relative ordering of the average transition energy
parameter $\Delta E$ for OO,OE and EE nuclei is the same as the one of
the GT$_+$ centroid energies of Ref.~\cite{Langanke00} and does not
seem to depend on the thermodynamic conditions.

In models (2) and (3) we assumed a linear isospin dependence of
$\Delta E$, but this choice has to be considered as the leading term
in an isospin dependence which is largely unknown.  Indeed the limited
$I$ range covered by present shell model calculations ($-0.07\leq I
\leq 0.22$) makes it impossible to pin down the most appropriate
functional trend. To demonstrate this statement, in model (4) we account 
for a quadratic dependence on isospin in addition to odd-even effects, see
appendix~\ref{app:fits}. The data are equally well reproduced by a
model with a linear (model(3)) or a quadratic (model(4)) isospin
dependence, see Figs.~\ref{fig:EC},\ref{fig:residuals} and
Table~\ref{table:chi2}.

To conclude, we have discussed here several physically realistic
effects~\cite{Koonin94} susceptible to improve the description of
microscopic EC rate calculations within a simple analytic
parameterization. A possible dependence on temperature, electron
density and isospin of the transition energy allow indeed for a 
better reproduction of data. Upon inclusion of odd-even effects, even
the staggering of individual rates due to fluctuations in the GT
strength distribution can be imitated. 

Since microscopic data are better reproduced, we expect that the
parameterizations discussed here provide, too, a better extrapolation
to regions where no large scale shell model calculation exist for the
moment.  It was in particular demonstrated in Ref.~\cite{Sullivan16}
that in simulations the maximum sensitivity on individual rates
concerns very neutron-rich nuclei around $N=50$.  These nuclei lie in
the lower right corner of Fig.\ref{fig:IQ}, with typical electron
capture $Q$-values between -17 and -12 MeV.  Those nuclei are
abundantly produced in the later stage of the collapse, for densities
$n_e \geq 1.4 \cdot 10^{-4}$ fm$^{-3}$ and temperatures $T \geq 1.2$
MeV~\cite{Sullivan16,Raduta16}.

We can see from Fig.\ref{fig:EC}, that the more refined
parameterizations, in particular models (2-4), predict an important
reduction of the EC rate for these nuclei with respect to the
reference parameterization from Ref.~\cite{Langanke03}.  To give a
representative example, at $n_e= 5.93 \cdot 10^{-5}$ fm$^{-3}$
and $T = 2.59$ MeV, model 3(4) predicts $\log_{10} (\lambda_{\mathrm{EC}})=1.69
(0.32)$ for the even-even $^{78}$Ni, $\log_{10} (\lambda_{\mathrm{EC}})=1.86 (0.70)$
for the even-odd $^{79}$Cu, and $\log_{10} (\lambda_{\mathrm{EC}})=2.49 (1.72)$ for
the odd-even $^{79}$Zn, to be compared with 
$\log_{10}(\lambda_{\mathrm{EC}}({\rm ^{78}Ni}))=3.11$, 
$\log_{10} (\lambda_{\mathrm{EC}}({\rm ^{79}Cu}))=3.34$ and 
$\log_{10}(\lambda_{\mathrm{EC}}({\rm ^{79}Zn}))=3.58$ obtained using a constant average value $\Delta
E=2.5$ MeV from Ref.\cite{Langanke03}. We therefore expect a
non-negligible influence on the global EC rate during core collapse,
see the discussion in the following section.

\section{NSE-averaged EC rates during core collapse}
\label{sec:average}
The effective reaction rate of a system composed of an ensemble of
nuclei, as it is the case of finite-temperature dilute nuclear matter
during core collapse, is the weighted sum of the contributions by individual nuclei: 
$$\langle \lambda_{\mathrm{EC}} \rangle =\sum_{A,Z} n(A,Z) \lambda_{\mathrm{EC}}(A,Z)/\sum_{A,Z}
n(A,Z)~.$$  
Depending on the
thermodynamic conditions, which affect both matter composition and
individual reaction rates, a few tens of nuclei dominate the
sum~\cite{Sullivan16}.

\begin{figure}
\begin{center}
\includegraphics[angle=0, width=0.99\columnwidth]{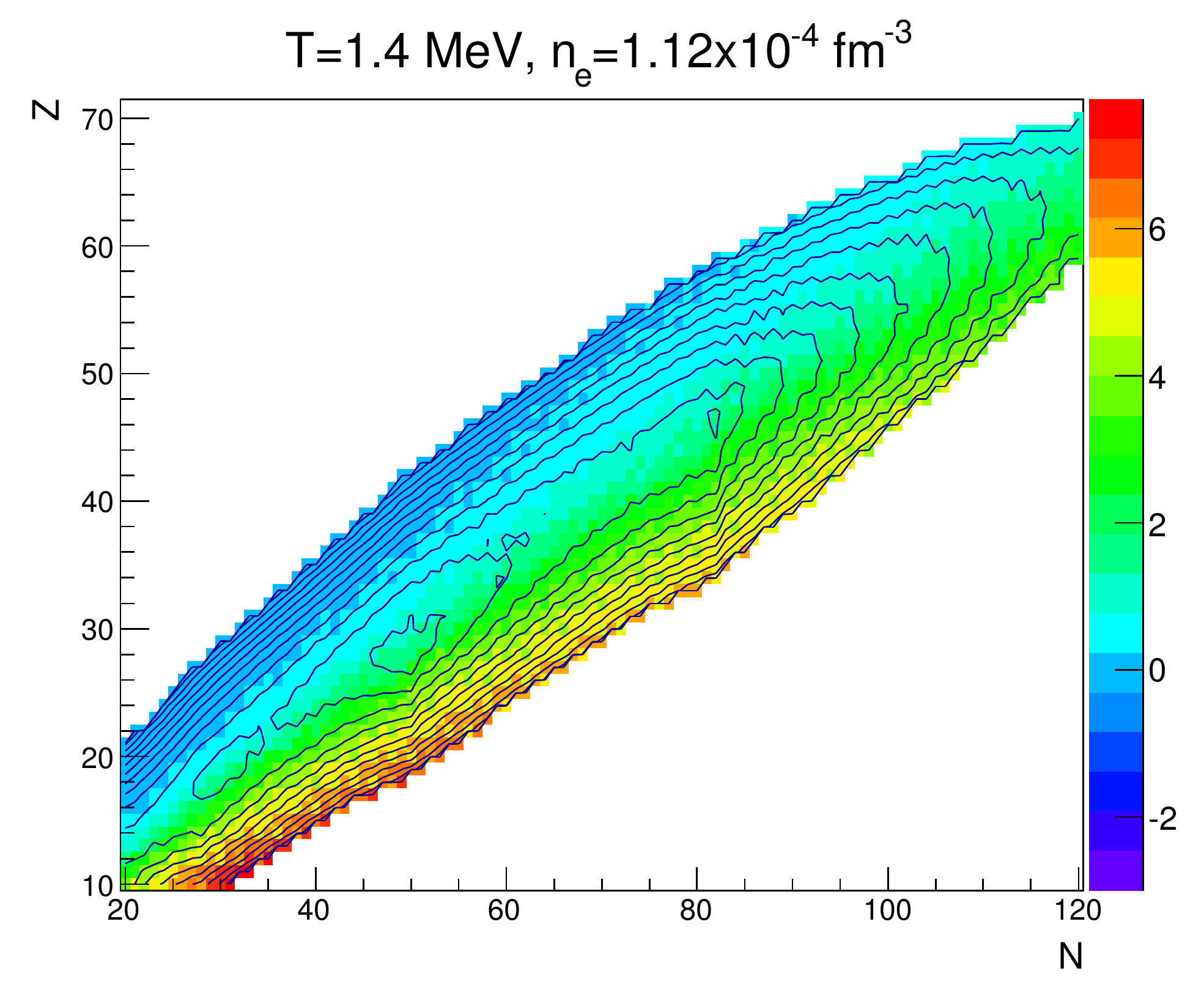}
\end{center}
\caption{(Color online) $T=1.4$ MeV, $n_B=3.52 \cdot 10^{-4}$ fm$^{-3}$, $Y_e=0.32$.
Ratio between individual EC rates corresponding to models (0) and (3) (color levels)
and isotopic multiplicities per unit volume (contours).
For both $\left(\lambda_{EC}^{(0)}/\lambda_{EC}^{(3)} \right)$ and $n(N,Z)$
$\log_{10}$-scales are employed.
\label{fig:indivEC}}
\end{figure}

As discussed in the previous section, the standard parameterization,
Eq.~(\ref{eq:lEC}), does not well describe EC on heavy neutron rich
nuclei. In particular, allowing for an isospin and odd-even dependence
may improve the reliability of the EC rates for those nuclei. To
give an example of the impact this might have on the collapse,
Fig. \ref{fig:indivEC} shows the ratio of EC
rates calculated according to models (0) and (3), 
across the nuclear chart for one
representative thermodynamic condition ($T=1.4$ MeV, $n_B=3.52 \cdot
10^{-4}$ fm$^{-3}$, $Y_e=0.32$).  In the same
figure, the contours give the isotopic abundances as predicted by
NSE. For both 
$\left( \lambda_\mathrm{EC}^{(0)}(A,Z)/\lambda_\mathrm{EC}^{(3)}(A,Z) \right)$ and $n(A,Z)$
$\log_{10}$-scales are used.
We can see that the improved parameterisation corresponds to
both higher and lower rates with respect to the reference model,
depending essentially on the isospin ratio. The most populated
nuclear species show rates 1-3 orders of magnitude lower than model
(0).

\begin{figure}
\begin{center}
\includegraphics[angle=0, width=0.9\columnwidth]{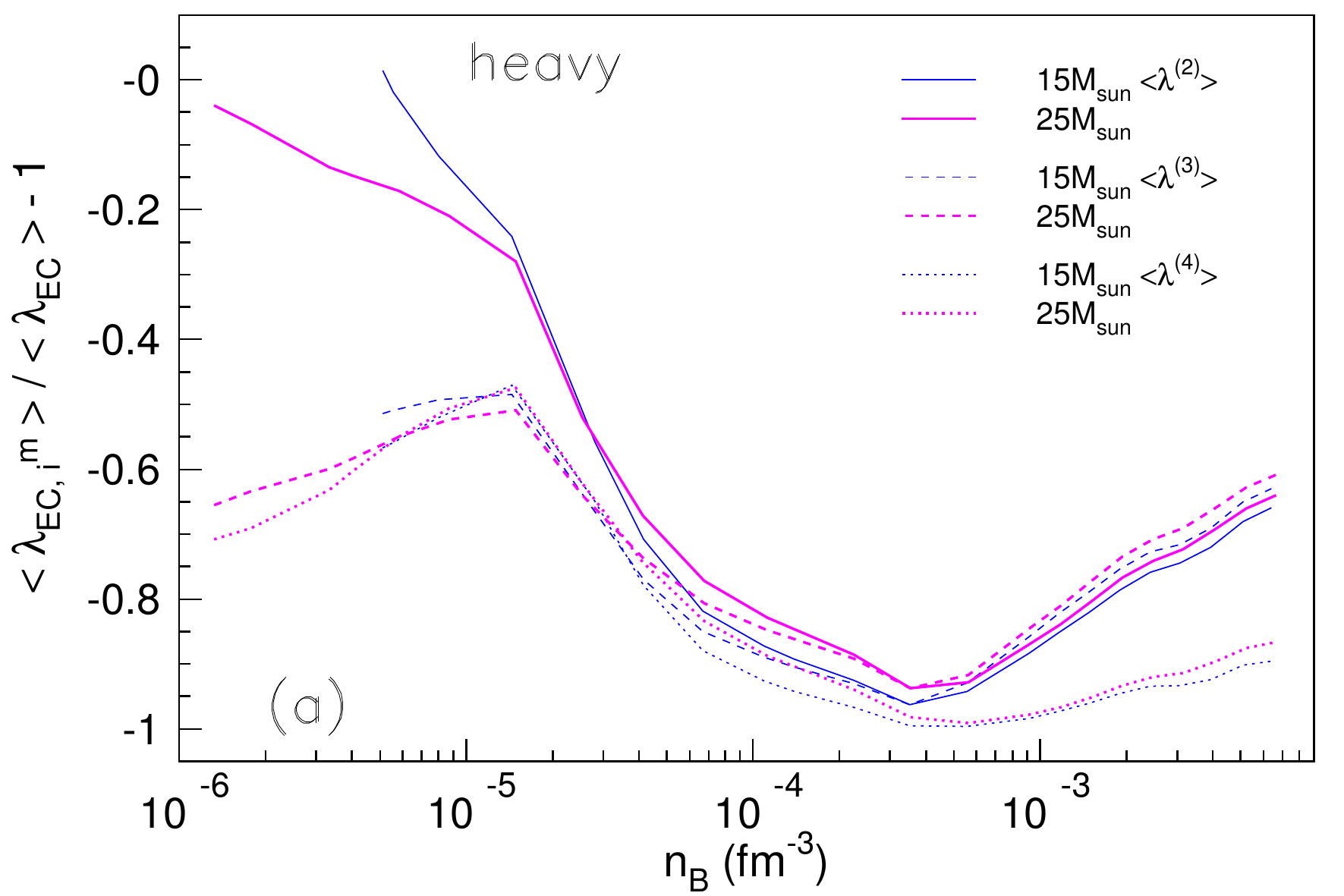}
\includegraphics[angle=0, width=0.9\columnwidth]{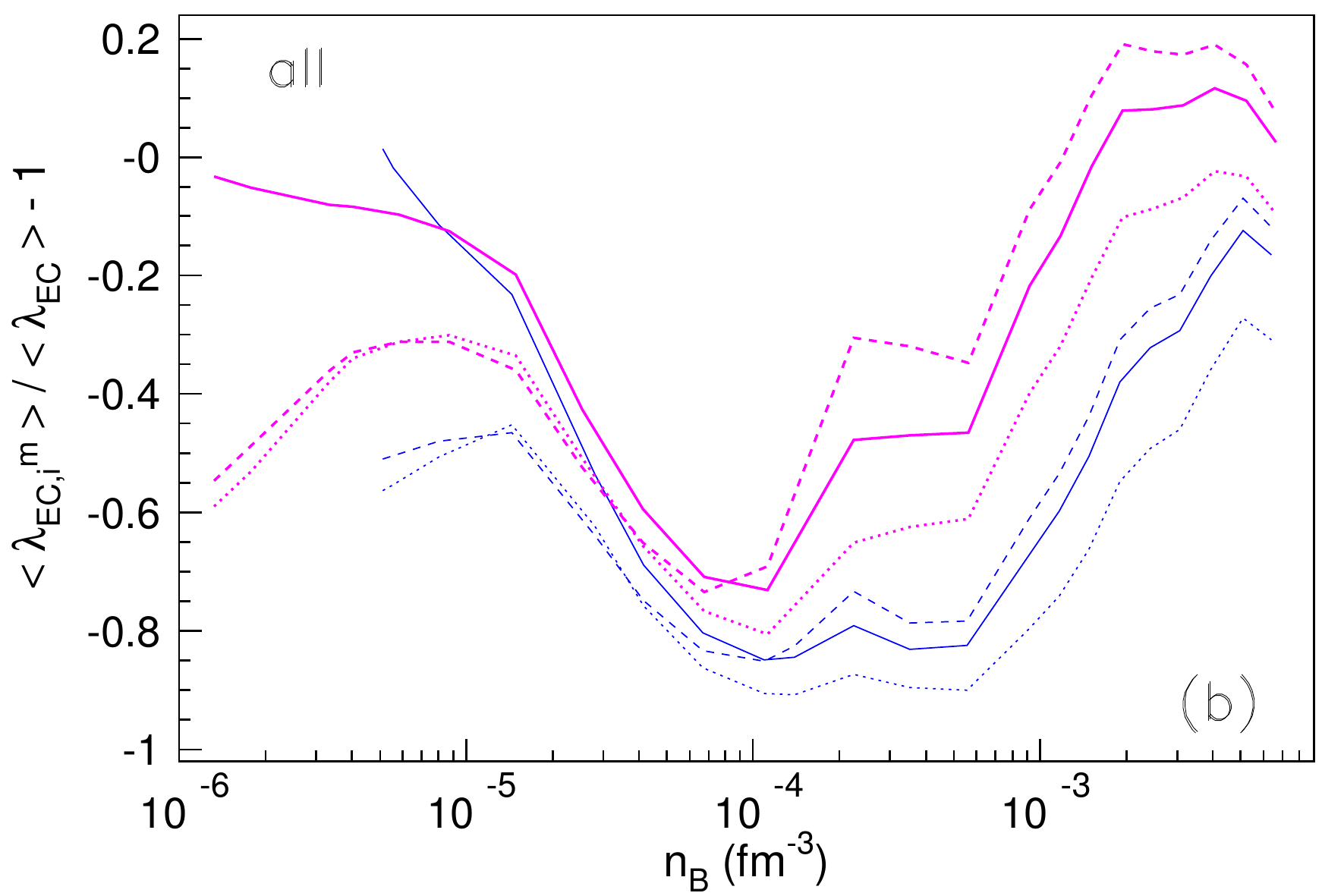}
\end{center}
\caption{Ratio between NSE-averaged EC rates using models (2), (3) and (4)
  for the effective transition energy parameter $\Delta E$ (see
  Sec.~\ref{sec:ecrates}) and the fiducial parameterization of
  Ref.~\cite{Langanke03}.  The thermodynamic conditions correspond to
  the central element during core collapse of two progenitors with $15
  M_{\odot}$ and $25 M_{\odot}$ \cite{Juodagalvis_2010}.  The average
  rate is calculated only on heavy nuclei ($A \geq 20$) in panel (a)
  and on all nuclei in panel (b).  
\label{fig:avEC}}
\end{figure}


The actual modification of the rates along the trajectories is reported in Fig.~\ref{fig:avEC}.
This figure shows the relative deviations in the NSE averaged EC rates of the improved
parameterizations, models (2), (3) and (4), with respect to the fiducial
model (0) from Ref. \cite{Langanke03} are depicted.  Contributions of all
(heavy) nuclei are considered separately.

The significant reduction of EC rates on heavy nuclei with large
negative-$Q$ values, accounted for by the isospin dependent
parameterizations, leads to a strong and almost systematic reduction
of the associated average EC rate, see panel (a).  In the first
stage of the collapse this reduction is independent of the choice
for the isospin dependence but depends strongly on whether odd-even
effects are included or not. At this stage, nuclei with $Q \lesssim
0$ showing strong nuclear structure effects dominate EC explaining
the sensitivity to odd-even effects. Isospin dependence is weak,
since the range of populated $Q$-values is well constrained by
microscopic data on $fp$-shell nuclei. 
In the late stages of the collapse electron capture reactions 
have processed the nuclear material into very neutron-rich
matter and low $Y_p \lesssim 0.33$ values are reached. 
That stage corresponds to densities above $n_B \approx 2 \cdot 10^{-4}$ fm$^{-3}$ 
(see fig. 3 of Ref. \cite{Raduta16}).
There, 
the isospin dependence dominates over structure details as can be seen from the similarity between the predictions of models (2) and (3). 
 The stronger isospin dependence in model (4) leads to a considerably stronger reduction of
the average EC rate on heavy nuclei with respect to model (3). 
As expected from Fig. \ref{fig:IQ}, which shows that
heavy nuclei produced in the central element of the two collapsing
cores populate the same $Q$-value domain, the NSE average EC rates on
heavy nuclei shows little sensitivity to the progenitor mass.

Panel (b) analyses EC rates NSE averaged over the whole nuclear
  distribution.  It shows that, as long as, due to low temperatures,
  protons and light nuclei have low multiplicities, heavy nuclei
  clearly dominate EC. This corresponds to baryon number densities
  below roughly $4 \cdot 10^{-5}$ fm$^{-3}$, where the mass fractions
  of protons and light nuclei do not exceed the percent level (see
  Ref. \cite{Raduta16}).  In the late stages of infall, when more
  protons and light nuclei are produced, a more moderate reduction of
  the overall rate is obtained.  The progenitor dependence of $\langle
  \lambda_{\mathrm{EC}}^{all} \rangle$ is due to the progenitor dependence in the
  matter composition. According to Ref. \cite{Raduta16} the mass
  fraction bound in heavy nuclei is higher in the core of the
  progenitor with the smaller mass.  Therefore higher deviations with
  respect to the fiducial model are obtained for the $15 M_{\odot}$
  progenitor than for the $25 M_{\odot}$ one.  

We recall that models (3) and (4) provide an equivalent fit to
the available microscopic calculations and only differ in their
extrapolation towards the extremely neutron-rich regime. This means
that the difference between the dashed and the dotted curves in
Fig.\ref{fig:avEC} can be considered as an estimate of the present
uncertainty due to unknown rates on extremely neutron rich nuclei.
However a word of caution is in order. Other factors, 
such as temperature-dependent Pauli-blocking and structure effects 
that might occur far from stability, could strongly influence the
strength distribution and depend on the region of nuclei investigated.
As a consequence, the true uncertainty might even be larger.
A correct extrapolation beyond the region covered by the shell-model
EC tables available in the literature can only be done by confronting
the predictions of our different phenomenological parameterizations to
reliable microscopic EC calculations for extremely neutron rich
nuclei, for instance with dedicated QRPA calculations
\cite{QRPA,inprep}.  If confirmed by microscopic calculations, this
reduction of the EC rate on neutron rich nuclei in hot and dense
environments would translate into higher electron fractions and
entropies of the inner core together with increased masses at bounce
and higher maxima of the neutrino luminosity peak~\cite{Sullivan16}.

We note that the reduction of the averaged EC rates observed in this paper
is opposite to the growth discussed in Ref.~\cite{Raduta16} due to an
expected magicity quenching towards the drip-line.  Which is the
dominant effect, which nuclei play the most important role and what
exactly are the consequences on a core collapse trajectory are
questions which can only be answered by core collapse simulations
treating consistently nuclear abundances, individual EC rates, and
neutrino dynamics. This task is beyond the present paper.

\section{Conclusions}
\label{sec:conclusions}
In this paper we have investigated the role played by EC on neutron
rich nuclei, abundantly produced in the late infall stages.
To this aim, we have considered thermodynamic conditions
corresponding to the trajectories of the central element, including a
mass of 0.05$M_{\odot}$, of two progenitors with 15$M_{\odot}$ and
25$M_{\odot}$ masses during collapse~\cite{Juodagalvis_2010}.
Modifications of EC rates are considered perturbatively in this
exploratory work, meaning that the time evolution of density,
temperature and electron fraction is not modified while different
prescriptions are considered for the EC rates.  

EC rates have been calculated according to an analytic
parameterization inspired by Ref.~\cite{Langanke03}. Specifically, we seek for an optimal
reproduction of large scale shell-model calculations considering a
possible temperature, density, and isospin dependence of the effective
average transition energy $\Delta E$.  Odd-even effects are also
included and shown to considerably improve the agreement with the
microscopic data. Under the considered thermodynamic conditions, these
new analytic parameterizations lead to an overall reduction of the
NSE-averaged EC rate of up to one order of magnitude.
This would imply, in an astrophysical simulation, larger electron
fractions and entropies in the inner core and a larger mass at bounce.
Though only an estimation due to the perturbative approximation of the
present approach, the observed reduction of the EC rate is important
enough to justify the effort for the calculation of accurate weak
interaction rates on extremely neutron rich nuclei in hot and dense
environments, constrained by new measurements of weak processes on
exotic nuclei.

\section*{Acknowledgments}
This work has been partially funded by NewCompstar, COST Action MP1304.  
Ad. R. R acknowledges kind hospitality from LPC-Caen and LUTH-Meudon.

\begin{appendix}
\section{Parameterizations of the electron capture rate}
\label{app:fits}

As mentioned in the text, we have employed four different
parameterizations of the electron capture rate. Within this appendix
we will detail the functional forms and the parameters employed. 

We made a least-square fit of the EC tables of Ref.~\cite{LMP} at each
grid point defined by a given value of temperature and electron
density (see Fig.\ref{fig:thermo}).  Since ${\mathcal B}$ gives only
an overall normalization and does not change the functional dependence
of the rate, we keep the fiducial value ${\mathcal B}=4.6$.
Conversely, we allow $\Delta E$ to depend on $T$, $n_e$ and on the
nuclear species, as it is physically reasonable to expect.

In model (1), the dependence on individual nuclei is neglected, 
but the average transition energy $\Delta E$ is assumed to
depend on the temperature and electron density. Thus, instead of employing a global value of $\Delta E$, we determine for each couple $T(i), n_e(j)$ given in the tables of the microscopic calculations of Ref.~\cite{LMP} a different value, 
\begin{equation}
\Delta E ^{(1)}(n_e(i),T(j))=a_{i,j} ~.
\end{equation}
The parameter $a_{ij}$ is thereby fitted to the rates at temperature
$T(i)$ and electron density $n_e(j)$.  

The values of the fit parameters $a_{ij}$ are listed in
Table~\ref{table:chi2} together with the corresponding
$\chi^2$-values.  The four considered temperatures and electron
densities have been chosen since they are the most relevant for the
CCSN trajectories we are interested in. They correspond to the highest
values of the tables of Ref.~\cite{LMP}.

For thermodynamic conditions lying in between the grid points,
$n_e(i)\leq n_e \leq n_e(i+1)$, $T(j) \leq T \leq T(j+1)$, a linear
dependence is assumed, \onecolumngrid
\begin{eqnarray}
\Delta E ^{(1)}(n_e,T)&=& 
T n_e \frac{a_{i+1,j+1}-a_{i,j+1}-a_{i+1,j}+a_{i,j}}
{\left[n_e(i+1)-n_e(i) \right]\left[T(j+1)-T(j) \right]} \nonumber \\
&+&n_e \frac{T(j+1) (a_{i,j+1}-a_{i,j})-T(j)(a_{i+1,j+1}-a_{i+1,j})}
{\left[n_e(i+1)-n_e(i) \right]\left[T(j+1)-T(j) \right]} 
+T \frac{n_e(i+1)(a_{i+1,j}-a_{i,j})-n_e(i)(a_{i+1,j+1}-a_{i,j+1})}
{\left[n_e(i+1)-n_e(i) \right]\left[T(j+1)-T(j) \right]} \nonumber \\
&+&
\frac{n_e(i+1) T(j+1) a_{ij}-n_e(i+1) T(j) a_{i+1,j}-n_e(i) T(j+1) a_{i,j+1}+n_e(i) T(j) a_{i+1,j+1}}
{\left[n_e(i+1)-n_e(i) \right]\left[T(j+1)-T(j) \right]}~.
\label{eq:interpol}
\end{eqnarray}
\twocolumngrid

In model (2) we allow for a possible $I=(N-Z)/A$-dependence of $\Delta E$ as:
\begin{equation}
\Delta E ^{(2)}(n_e(i),T(j),I)=b_{i,j}I +c_{i,j} 
\label{eq:model2}
\end{equation}
Again as in model $(1)$, we assume a linear evolution of $\Delta E$
between different points $T(i),n_e(j)$ according to
Eq. (\ref{eq:interpol}), upon replacing $a_{i,j}=b_{i,j}I+c_{i,j}$.
Columns 8-10 in Table~\ref{table:chi2} provide, for the same
temperature and electron density conditions as before, the $\chi^2$
values corresponding to this second scenario together with the values
of $b_{i,j}$ and $c_{i,j}$.  As expected, the extra degree of freedom
leads to a better fit for all temperatures and/or densities expressed
by lower $\chi^2$ values. The still relatively high values for
$\chi^2$ obtained at the lowest considered temperatures arise due to
fluctuations of individual rates, more than due to global trends in the
Gamow-Teller phenomenology.

In model (3) we make a separate fit of the average transition energy
parameter $\Delta E$ for odd-odd (OO), even-even (EE) and odd-even
(OE) nuclei:
\begin{eqnarray}
\Delta E ^{(3)}_{OO}(n_e(i),T(j),I)&=&b_{i,j}^{OO}I +c_{i,j}^{OO}, \label{eq:31}\\ 
\Delta E ^{(3)}_{OE}(n_e(i),T(j),I)&=&b_{i,j}^{OE}I +c_{i,j}^{OE}, \label{eq:32}\\ 
\Delta E ^{(3)}_{EE}(n_e(i),T(j),I)&=&b_{i,j}^{EE}I +c_{i,j}^{EE}~.  \label{eq:33}
\end{eqnarray}
As we can see from Table~\ref{table:chi2} (columns 11-13), the fit is
considerably improved and very reasonable $\chi^2$ are obtained for
all thermodynamic conditions considered here.

The isospin dependence of $\Delta E(I)$ is largely unknown, and the
linear choice of model (2) and model (3), see
Eqs.(\ref{eq:model2}-\ref{eq:33}), has to be considered as the leading
term. In model (4) we introduce a quadratic isospin dependence,
\begin{eqnarray}
\Delta E ^{(4)}_{OO}(n_e(i),T(j),I)&=&b_{i,j}^{OO}I^2 +c_{i,j}^{OO}, \label{eq:41}\\ 
\Delta E ^{(4)}_{OE}(n_e(i),T(j),I)&=&b_{i,j}^{OE}I^2 +c_{i,j}^{OE}, \label{eq:42}\\ 
\Delta E ^{(4)}_{EE}(n_e(i),T(j),I)&=&b_{i,j}^{EE}I^2 +c_{i,j}^{EE}  \label{eq:43}
\end{eqnarray}
We can see from Table ~\ref{table:chi2} (columns 14-16) that linear
and quadratic $I$-dependence of $\Delta E$ lead to similar $\chi^2$
values.

\begin{table*}[t]
\centering \scalebox{0.82}{
\begin{tabular}{ccc|cc|cc|ccc|ccc|ccc}
\hline
$T$ & $n_e$ & $\mu_e$ & \multicolumn{2}{c}{$\Delta E^{(0)}$} & \multicolumn{2}{c}{$\Delta E^{(1)}$} & \multicolumn{3}{c}{$\Delta E^{(2)}$} & \multicolumn{3}{c}{$\Delta E^{(3)}$}& \multicolumn{3}{c}{$\Delta E^{(4)}$} \\
(MeV) & (fm$^{-3}$) & (MeV) & $\Delta E$ & $\chi^2$  & $\Delta E$ & $\chi^2$  & $b$ & $c$ & $\chi^2$ & $b$ & $c$ & 
$\chi^2$ & $b$ & $c$ & $\chi^2$ \\
\hline
0.43 & $5.93 \times 10^{-8}$ & 2.18 & 2.5 & $1.22 \times 10^{2}$ &  1.77 & $9.68 \times 10^{1}$ & 
 $-1.42\times 10^1$&  3.31 & $7.74\times 10^1$ &
$-2.01\times 10^1$ & 4.11&  $1.06\times 10^1$ & $-9.63 \times 10^1$&  3.23 & $1.32 \times 10^1$ \\
     &                     &                 &      &   &&       &      &       &       &    
$-2.97 \times 10^1$ & 6.69 & 5.83 & $-1.38 \times 10^2$&  5.35 &  6.42 \\ 
     &                     &                 &      &   &&       &      &       &       &
$ -1.67 \times 10^1$ & 2.25  &1.89 & $-8.24 \times 10^1$ & 1.58 & 2.94 \\
\hline
0.43 & $5.93 \times 10^{-7}$ & 5.03 & 2.5 & $1.05\times 10^{2}$ & 1.86 & $9.01\times 10^{1}$ &   
$-1.37\times 10^1$ & 3.50 & $7.95\times 10^1$ &
$-2.26\times 10^1$ & 4.85 & $1.64\times 10^1$ &  $-1.04 \times 10^2$ & 3.77 & $1.74 \times 10^1$ \\
     &                     &                 &      &   &&       &      &       &       &  
$-3.46 \times 10^1$ & 7.83 & $1.00 \times 10^1$&  $-1.62 \times 10^2$&  6.30 & 9.14 \\    
 &                     &                 &      &   &&       &      &       &       & 
$-1.78 \times 10^1$ & 2.66 & 3.38 &  $-8.28 \times 10^1$  & 1.83  & 3.88\\
\hline
0.43 & $5.93 \times 10^{-6}$ & $1.10 \times 10^{1}$& 2.5 & $4.00\times 10^{1}$ & 2.26 & $3.93\times 10^{1}$  &
$-8.26$  & 3.37  & $3.85\times 10^1$ & 
 $-1.07\times 10^1$ & 4.34 & $1.22\times 10^1$ &  $-6.00 \times 10^1$ & 4.05 & $1.20 \times 10^1$\\
     &                     &                 &      &   &&       &      &       &        &
 $-2.26 \times 10^1$ & 7.54 & 6.18 &  $-1.16 \times 10^2$ & 6.92 & 5.35\\
     &                     &                 &      &   &&       &      &       &       &  
 $-9.58$ & 2.40 & 3.38 & $-5.01 \times 10^1$ & 2.09 & 3.22 \\
\hline
0.43 & $5.93 \times 10^{-5}$ & $2.38 \times 10^{1}$& 2.5 & $1.25\times 10^{1}$ & 4.01 & $1.15\times 10^{1}$& 
5.28 $\times 10^{1}$ & $-1.87$ &  7.64 & 
 $5.40\times 10^1$&  $-1.70$ & 3.68 &  $2.46 \times 10^2$ & $6.62 \times 10^{-1}$ & 4.12\\
     &                     &                 &      &   &&       &      &       &       & 
 $4.53 \times 10^1$ & $7.90 \times 10^{-1}$ & 1.70 & $1.77 \times 10^2$ & 3.24 & 1.95 \\ 
     &                     &                 &      &   &&       &      &       &       &  
 $5.21 \times 10^1$ & $-3.16$ & 1.42 & $2.29 \times 10^2$ & $-8.12 \times 10^{-1}$&  1.65 \\
\hline
0.86 & $5.93 \times 10^{-8}$ & 1.48 & 2.5 & $3.54\times 10^{1}$& 2.6 & $3.53\times 10^{1}$ & 
$-1.04 \times 10^{1}$ & 3.62 &  $3.19 \times 10^{1}$ &
$-1.50\times 10^1$ & 4.22 & 3.51 &  $-8.61 \times 10^1$ & 3.77&  3.44\\
  &                     &                 &      &   &&       &      &       &       &  
  $-1.86 \times 10^1$ & 6.52 &  2.75 & $-1.05 \times 10^2$&  5.97 & 2.18  \\
  &                     &                 &      &   &&       &      &       &       &  
 $-1.37 \times 10^1$ & 2.06 & $8.01 \times 10^{-1}$ & $-7.55 \times 10^1$&  1.61 & $8.83 \times 10^{-1}$ \\
\hline
0.86 & $5.93 \times 10^{-7}$ & 4.68 & 2.5 & $3.28\times 10^{1}$ & 2.67 & $3.25\times 10^{1}$ &   
$-8.19$ &  3.56 & 3.10 $\times 10^{1}$ &
 $-1.46\times 10^1$&  4.47 & 5.05& $-8.29 \times 10^1$&  4.01 &  4.69 \\
  &                     &                 &      &   &&       &      &       &       &
$ -2.00 \times 10^1$ & 6.99 & 3.85 & $-1.12 \times 10^2$ & 6.41 & 3.16\\
  &                     &                 &      &   &&       &      &       &       & 
$ -1.28 \times 10^1$ & 2.24 & 1.16 &$-7.01 \times 10^1$&  1.79 & 1.11 \\ 
\hline
0.86 & $5.93 \times 10^{-6}$ &$1.08 \times 10^{1}$& 2.5 & $2.18\times 10^{1}$ & 2.96 & $2.08 \times 10^{1}$     & 
1.94 & 2.72&  2.08 $\times 10^{1}$ & 
 $-1.52$ & 3.56 & 6.49 & $-2.07 \times 10^1$ & 3.72 & 6.43\\
     &                     &                 &      &   &&       &      &       &       &   
 $-1.06 \times 10^1$ & 6.39  &3.71 & $-7.00 \times 10^1$ & 6.37 & 3.43 \\
     &                     &                 &      &   &&      &      &       &       & 
 $-1.76$ & 1.60 & 1.90 & $-2.07 \times 10^1$&  1.73 & 1.87 \\
\hline
0.86 & $5.93 \times 10^{-5}$ &$2.37 \times 10^{1}$&2.5 & $1.19\times 10^{1}$ & 3.85 & $1.12\times 10^{1}$      &  
$5.57 \times 10^{1}$ & $-2.25$ &  7.20 & 
$5.64\times 10^1$ & $-2.03$ & 3.45 &$2.59 \times 10^2$ & $3.95 \times 10^{-1}$ & 3.89 \\
     &                     &                 &    &    &&      &      &       &        &  
 $4.85 \times 10^1$ & $4.52 \times 10^{-1}$ & 1.63 &$1.93 \times 10^2$ & 3.00 & 1.89 \\
     &                     &                 &    &    &&      &      &       &        &  
$5.37 \times 10^1$ & $-3.54$ & 1.25 & $2.39 \times 10^2$  &$-1.16$ & 1.47 \\
\hline
2.59 & $5.93 \times 10^{-8}$ & $2.05 \times 10^{-1}$& 2.5 & 7.25 & 2.13 & 7.11 & $1.16 \times 10^{1}$ & 1.10 & 6.67 &  8.09 & 1.46 &  1.52 & 
$3.02 \times 10^1$ & 1.85 & 1.59 \\
     &                     &                 &      &   &&       &      &       &       & $1.34 \times 10^1$ & 3.42 & 1.48 & $4.83 \times 10^1$ & 4.14 & 1.56 \\
     &                     &                 &      &   &&       &      &       &       & 4.62 & $-6.65 \times 10^{-1}$ &$ 5.33 \times 10^{-1}$ &$1.71 \times 10^1$ & $-4.44 \times 10^{-1}$ & $5.43 \times 10^{-1}$ \\
\hline
2.59 & $5.93 \times 10^{-7}$ & 1.94 & 2.5 & 7.36 & 2.13 & 7.22 & $1.35 \times 10^{1}$ & $8.94 \times 10^{-1}$ & 6.69  & 9.61 & 1.31 & 1.61 &$3.34 \times 10^1$ & 1.82 & 1.70 \\
    &                     &                 &      &   &&       &      &       &       &$ 1.39 \times 10^1$  &3.37 & 1.50 &  $4.85 \times 10^1 $& 4.16 &  1.59 \\ 
    &                     &                 &      &   &&       &      &       &       & 5.83 &  $-7.65 \times 10^{-1}$ & $5.58 \times 10^{-1}$ &$2.07 \times 10^1$ & $-4.67 \times 10^{-1}$ & $5.72 \times 10^{-1}$  \\ 
\hline
2.59 & $5.93 \times 10^{-6}$ &  9.09 & 2.5 & 7.70 & 2.36 & 7.69 & $2.63 \times 10^{1}$ & $-2.90 \times 10^{-1}$ & 6.27 &
$2.24\times 10^1$ & $2.09 \times 10^{-1}$ & 1.94&  $9.49 \times 10^1$ & 1.27 & 2.16 \\
     &                     &                 &     &    &&      &      &       &    &   $2.34 \times 10^1$ & 2.45 & 1.54 &
$8.92 \times 10^1$ & 3.72 & 1.68 \\
     &                     &                 &     &    &&     &      &       &    & $1.91 \times 10^1$ & $-1.74$ &  $6.75 \times 10^{-1}$  &$8.35 \times 10^1$ & $-8.86 \times 10^{-1}$&  $7.44 \times 10^{-1}$ \\
\hline
2.59 & $5.93 \times 10^{-5}$ &$2.29 \times 10^{1}$ & 2.5 & 9.08 & 2.50 & 9.08 & $7.24 \times 10^{1}$ & -4.75 &  5.07 &
$ 6.92\times 10^1$ & $-4.34$ & 2.26 &$3.34 \times 10^2$&  -1.65&  2.66\\
     &                     &                 &     &    &&      &      &       &     &   $6.82 \times 10^1$ & $-2.10$ & 1.33 &
 $2.95 \times 10^2$ & 1.02&  1.58\\
     &                     &                 &     &    &&      &      &       &     &  $6.71 \times 10^1$ & $-5.99$ & $8.23 \times 10^{-1}$ &$3.23 \times 10^2$ & $-3.41$ & $9.93 \times 10^{-1}$ \\
\hline
8.62 & $5.93 \times 10^{-8}$ & $1.84 \times 10^{-2}$& 2.5 & 8.17 &  -1.87 &  6.85 & $1.01 \times 10^{2}$ &  $-9.99$ & 4.51 &
$ 9.17\times 10^1$ & $-9.16$ & 2.17&
$5.05 \times 10^2$ & $-6.72$ & 2.43\\
     &                     &                 &     &    &&      &      &       &     &  $1.23 \times 10^2$  &$-9.03$ & 1.12 & 
$6.16 \times 10^2$ & $-5.00$ & 1.30 \\
    &                     &                 &     &    &&      &      &       &     & $7.26 \times 10^1$ & $-1.06 \times 10^1$ &  $7.64 \times 10^{-1}$ & $3.72 \times 10^2$ & $-8.44$ &$ 8.60 \times 10^{-1}$ \\
\hline
8.62 & $5.93 \times 10^{-7}$ & $1.84 \times 10^{-1}$& 2.5 & 8.17 & $-1.86$ &  6.85 & $1.01 \times 10^{2}$ & $-1.00 \times 10^{1}$ & 4.50  & $9.19\times 10^1$ & $-9.18$ & 2.17 &
$5.05 \times 10^2$  &$-6.72$&  2.43 \\
     &                     &                 &     &    &&      &      &       &      & $1.23 \times 10^2$ & $-9.05$  &1.12 &
 $6.16 \times 10^2$ & $-5.00$&  1.30\\
     &                     &                 &     &    &&      &      &       &     & $ 7.27 \times 10^1$ & $-1.06 \times 10^1$ & $7.62 \times 10^{-1}$ &$3.72 \times 10^2$ & $-8.44$ & $8.58 \times 10^{-1}$ \\
\hline
8.62 & $5.93 \times 10^{-6}$ & 1.83 & 2.5 & 8.16 & $-1.83$ &  6.87 & $1.03 \times 10^{2}$ & $-1.02 \times 10^{1}$ & 4.52 & 
$9.35\times 10^1$ & $-9.35$ & 2.17 &
$5.08 \times 10^2$ & $-6.75$&  2.45 \\
     &                     &                 &     &    &&      &      &       &     & $1.24 \times 10^2$ & $-9.20$ & 1.12 &
$6.17 \times 10^2$ & $-5.02$ & 1.31 \\ 
     &                     &                 &     &    &&      &      &       &     & $7.39 \times 10^1$ &  $-1.07 \times 10^1$ & $7.67 \times 10^{-1}$ &$3.75 \times 10^2$ & $-8.46$ & $8.65 \times 10^{-1}$ \\ 
\hline
8.62 & $5.93 \times 10^{-5}$ & $1.44 \times 10^{1}$ & 2.5 & 8.43 & $-2.17$ & 7.21 & $1.23 \times 10^{2}$ & $-1.25 \times 10^{1}$& 4.55 & 
$1.14 \times 10^2$  &$-1.16\times 10^1$ & 2.22 &
$5.93 \times 10^2$&  $-8.04$ & 2.53\\
     &                     &                 &     &    &&      &      &       &     & $1.43 \times 10^2$ & $-1.14 \times 10^1$ & 1.13 &
$6.86 \times 10^2$ & $-6.19$ & 1.34\\
     &                     &                 &     &    &&      &      &       &     &  $9.13 \times 10^1$ & $-1.26 \times 10^1$ & $7.87 \times 10^{-1}$ &$4.50 \times 10^2$ & $-9.61$  &$9.02 \times 10^{-1}$ \\
\hline
\end{tabular}
}
\caption{Values of $\Delta E$-parameter (MeV) of Eq.~(\ref{eq:lEC})
  extracted by fitting shell model EC rates of Ref.~\cite{LMP} under
  different hypothesis (see text) and for various thermodynamic
  conditions mentioned in columns 1 and 2.  In case of models $(3)$
  and $(4)$, values of odd-even, odd-odd and even-even parameters are
  given respectively in rows 1, 2, 3. $\mu_e$ represents the electron
  chemical potential and includes the rest mass.
\label{table:chi2}}
\end{table*}

\end{appendix}

\end{document}